\documentclass[aps,pra,twocolumn,showpacs,superscriptaddress]{revtex4}
\usepackage{graphicx}
\usepackage{amsmath}
\usepackage{amssymb}

\input{epsf}
\begin{document}

\title{Trapped two-component
Fermi gases with up to six particles:
Energetics, structural properties,
and molecular condensate fraction}

\author{D. Blume and K. M. Daily}
\affiliation{Department of Physics and Astronomy,
Washington State University,
  Pullman, Washington 99164-2814, USA}

\date{\today}

\begin{abstract}
We investigate
small equal-mass two-component Fermi gases under external
spherically symmetric confinement in which atoms with opposite
spins interact through a short-range two-body model potential.
We employ a non-perturbative microscopic framework,
the stochastic
variational approach, and
determine the system properties as functions of the
interspecies $s$-wave scattering length
$a_s$, the orbital angular 
momentum $L$ of the
system, and the numbers $N_1$ and $N_2$ of spin-up and spin-down atoms
(with $N_1-N_2 =0$ or 1 and $N \le 6$, where $N=N_1+N_2$). 
At unitarity, we determine the energies
of the five- and six-particle systems
for various ranges $r_0$ of the underlying two-body model
potential and extrapolate to the zero-range limit. These energies 
serve as benchmark results that can be used to validate 
and assess other numerical approaches.
We also present
structural properties
such as the pair distribution function and the radial density.
Furthermore, 
we analyze the one-body and two-body density matrices.
A measure for the molecular condensate fraction 
is proposed and applied.
Our calculations show explicitly that the 
natural orbitals and the momentum distributions
of atomic Fermi gases approach those characteristic for 
a molecular
Bose gas if the $s$-wave scattering length $a_s$, $a_s>0$,
is sufficiently small.
\end{abstract}


\maketitle

\section{Introduction}
Over the past few years, the interest in small trapped Bose and Fermi gases, and mixtures 
thereof,
has increased tremendously
for a number of reasons. 
First, atomic gases provide
an ideal platform for investigating phenomena related to Efimov 
physics~\cite{efim71,efim73,braa06}.
While the majority of
investigations of the Efimov effect have focused on the three-body
system, larger systems
have attracted considerable attention recently
from theoretical and experimental 
groups~\cite{plat04,yama06,hann06,hamm07,stec09a,wang09,stec10,yama10,cast10,ferl09,zacc09,poll09}.
Second,
small trapped atomic systems can be realized by loading an atomic gas
into an optical lattice~\cite{grei02,koeh05,thal06,bloc08}. 
If the tunneling between lattice
sites is small
and if the interactions between neighboring sites can be neglected,
then each lattice site provides a realization
of a trapped few-body system.
In this setting, one interesting prediction is that
effective three- and higher-body interactions should 
emerge~\cite{johnsonNJP}.
Third,
small atomic gases can be viewed
as a bridge between two-body and many-body 
systems (see, e.g., Refs.~\cite{blum07,stec08,chan07,bulg07}). 
In most cases, the
two-body system is well characterized, making a bottom-up approach  attractive.
Such an approach
treats increasingly larger
systems and eventually connects observables
for mesoscopic systems
 with
those predicted by many-body theories, e.g., through the use of the
local density approximation.
Fourth, few-body systems often times allow for highly accurate
treatments, thereby providing much needed benchmark results.
For example,
a number of 
lattice-based approaches are presently being applied to 
trapped cold atom 
systems (see Refs.~\cite{chen04,bulg06,buro06,lee06,lee06a,abe09} 
for lattice-based treatments of
the homogeneous system).
While these approaches promise to be
very powerful, currently only a few benchmark
results are available that allow for a careful 
assessment of their validity regimes.

This paper treats equal-mass
two-component Fermi gases under external harmonic confinement
with short-range $s$-wave interactions.
Our work builds on the rapidly expanding number of papers
that treat trapped three-dimensional few-fermion 
systems (see, e.g., Refs.~\cite{blum07,stec08,chan07,bulg07,cast04,wern06,wern06a,kest07,stet07,stec07c,stec07b,alha08,blumpolarized,blum09a,liu09,dail10}).
The ground state of trapped equal-mass two-component Fermi gases, e.g., has
been investigated numerically by the fixed-node
diffusion Monte Carlo 
approach~\cite{blum07,stec08,chan07,blumpolarized} and the
stochastic variational approach~\cite{stec07c,blum07,stec08,blum09a,dail10}.
In the strongly-interacting unitary regime,
the properties of the 
system---motivated by analytical treatments 
that exploit the scale invariance of equal-mass Fermi
gases at unitarity~\cite{cast04,wern06}---have been
interpreted
within the
hyperspherical framework~\cite{blum07,stec08}.
In some cases, the excitation spectrum at unitarity 
has also been 
investigated~\cite{blum07,stec08,wern06,wern06a}.
In addition, small two-component Fermi gases have been investigated
as a function of the $s$-wave scattering length 
$a_s$~\cite{kest07,stec07c,stet07,stec07b,stec08,liu09,blum09a,dail10}. 
For small $|a_s|$, $a_s<0$,
the energy crossover
curve has been analyzed by applying first order perturbation theory
to a weakly-attractive atomic Fermi gas~\cite{stec07b,stec08,dail10}.
For small $|a_s|$, $a_s>0$, in contrast, 
the energy crossover curve 
has been analyzed by applying first order perturbation theory
to a weakly-repulsive molecular gas~\cite{stec07b,stec08,dail10} 
(see also Refs.~\cite{astr04c,petr04aa,petr05}).
Small two-component Fermi gases have also provided the first high
precision tests~\cite{blum09a} 
of the Tan relations~\cite{tan08a,tan08b,tan08c} that apply to both 
inhomogeneous and homogeneous $s$-wave interacting Fermi gases.

Following up on our earlier work, this paper 
presents 
new results for trapped equal-mass Fermi gases
with up to $N=6$, where $N=N_1+N_2$
and $N_1-N_2=0$ or 1.
Our main results are:
(i) We report extrapolated zero-range energies for five- and 
six-particle systems 
with $(N_1,N_2)=(3,2)$ and $(3,3)$ for
various angular momenta at unitarity.
(ii) We present energy crossover curves for the $(N_1,N_2)=(3,2)$
system
for the ground state and various excited states.
(iii) We present a detailed analysis of the dependence
of the few-particle energies on the range of the underlying two-body
potential.
(iv) We present structural properties for the 
$(N_1,N_2)=(2,1)$, $(2,2)$, $(3,2)$ and $(3,3)$ systems throughout the
crossover, including unitarity.
(v) We quantify the correlations of 
few-fermion systems by analyzing the one- and two-body density
matrices as well as the 
momentum distributions.
In particular, we propose a measure of the 
molecular condensate fraction
and apply it to few-fermion systems with up to $N=6$ atoms.
Related analyses have previously been pursued for 
bosonic gases~\cite{dubo01,mous02,thog07} and one-dimensional 
systems~\cite{gira00,deur07,casu08}, 
but we are not aware of analogous studies for
trapped three-dimensional two-component Fermi gases.

Section~\ref{sec_theory} introduces the system Hamiltonian
and the stochastic variational approach employed to solve the
time-independent Schr\"odinger equation
for small trapped two-component systems.
In addition, Sec.~\ref{sec_theory} reviews the 
definitions of the one- and
two-body density matrices 
and their relationship to the natural orbitals and momentum
distribution.
Section~\ref{sec_results} presents and interprets 
our results for 
various parameter combinations.
Lastly, Section~\ref{sec_conclusions} summarizes
our main results and concludes.
Mathematical derivations and discussions of technical aspects are 
collected in
Appendices~\ref{sec_appendixa} through \ref{sec_appendixc}.

\section{Theoretical background}
\label{sec_theory}

\subsection{System Hamiltonian}
\label{sec_hamiltonian}
Our model Hamiltonian that describes
equal-mass two-component Fermi gases with
$N_1$ spin-up and $N_2$ spin-down atoms ($N=N_1+N_2$
and $N_1 \ge N_2$)
under external spherically symmetric 
harmonic
confinement with angular trapping frequency $\omega$
reads
\begin{eqnarray}
\label{eq_ham}
H= 
\sum_{j=1}^{N} \left( \frac{-\hbar^2}{2 m_a} \nabla_{\vec{r}_j}^2 +
\frac{1}{2}m_a \omega^2 \vec{r}_j^2 \right) + \nonumber \\
\sum_{j=1}^{N_1} \sum_{k=N_1+1}^N V_{\mathrm{tb}}(r_{jk}).
\end{eqnarray}
Here, $m_a$ denotes the atom mass and
$\vec{r}_j$
the position vector of 
the $j$th particle measured with respect to the trap
center (with $r_{jk}=| \vec{r}_j-\vec{r}_k|$);
the first $N_1$ position vectors correspond to the spin-up
atoms and the last $N_2$ position vectors to the
spin-down atoms.
Hamiltonian~(\ref{eq_ham}) assumes that like fermions
are non-interacting.
The interspecies interactions are
modeled through a purely attractive
Gaussian two-body  potential 
$V_{\mathrm{tb}}(r)$, 
\begin{eqnarray}
\label{eq_fr}
V_{\mathrm{tb}}(r)=-V_0 \exp \left[-\left( \frac{r}{\sqrt{2}r_0} \right)^2
\right].
\end{eqnarray}
We take the range $r_0$ to be much smaller than
the harmonic oscillator length $a_{\mathrm{ho}}$, where 
$a_{\mathrm{ho}}=\sqrt{\hbar/(m_a \omega)}$.
The depth $V_0$, $V_0>0$, and the range $r_0$ are adjusted so 
that the free-space two-body $s$-wave scattering length $a_s$ 
takes on the desired value. 
We restrict ourselves to two-body
potentials that support no free-space
$s$-wave two-body bound state and  one free-space $s$-wave
two-body bound state for negative $a_s$ and positive $a_s$, respectively.
If the scattering length $a_s$ is notably larger than the
range $r_0$, then the
properties of
small trapped two-component Fermi gases are universal, i.e.,
independent of the details of the underlying two-body 
potential~\cite{gior08,astr04c,bake99,ohar02,ho04,tan04,chan05,chan04a,thom05,son06,stew06}.
Thus, we limit ourselves to parameter combinations with $r_0 \ll a_s$ 
and $r_0 \ll a_{\mathrm{ho}}$. For these parameter combinations,
energy shifts due to $p$-wave or higher partial wave scattering between unlike
fermions
are negligible.
In a few cases, we perform calculations for different
$r_0$ and explicitly extrapolate to the $r_0 \rightarrow 0$
limit. 

Our goal is to solve the time-independent Schr\"odinger
equation for the Hamiltonian given in Eq.~(\ref{eq_ham}),
and to analyze the energy spectrum and structural properties.
To this end, we use that the total wave function
$\psi_{\mathrm{tot}}(\vec{r}_1,\cdots,\vec{r}_N)$
separates into a relative part $\psi_{\mathrm{rel}}$ 
and a center-of-mass part $\psi_{\mathrm{cm}}$.
The relative wave function $\psi_{\mathrm{rel}}$ is written in terms 
of Jacobi vectors $\vec{\rho}_1,\cdots,\vec{\rho}_{N-1}$;
its determination through the 
stochastic variational approach is reviewed briefly in the next subsection.
Throughout, we assume that center-of-mass excitations are absent,
i.e., we assume that the center of mass wave function
is given by
\begin{eqnarray}
\label{eq_cm}
\psi_{\mathrm{cm}}(\vec{R}_{\mathrm{cm}})=
N_{\mathrm{cm}}\exp \left( -\frac{\vec{R}_{\mathrm{cm}}^2}{2 a_{\mathrm{ho}}^2/N} \right),
\end{eqnarray}
where $N_{\mathrm{cm}}$ denotes a normalization constant
and $\vec{R}_{\mathrm{cm}}$ the center of mass vector, 
$\vec{R}_{\mathrm{cm}}=\sum_{j=1}^N \vec{r}_j/N$.
The relative wave function $\psi_{\mathrm{rel}}$
is a simultaneous eigen function 
of the relative Hamiltonian $H_{\mathrm{rel}}$,
the square of the relative orbital angular
momentum operator, the $z$-projection of the relative orbital
angular momentum operator and the parity operator.
Correspondingly, $\psi_{\mathrm{rel}}$ and the associated 
eigen energies $E_{\mathrm{rel}}$ are labeled
by the quantum numbers $L$, $M_L$ and $\Pi$.

\subsection{Stochastic variational treatment}
\label{sec_sv}
To determine the relative eigen functions $\psi_{\mathrm{rel}}$ and 
relative eigen energies $E_{\mathrm{rel}}$,
we employ the stochastic variational (SV) 
approach~\cite{varg95,varg01,cgbook,sore05}.
Our implementation follows that described in 
Refs.~\cite{stec07c,stec08,dail10},
and here we only emphasize a few key points.
The SV approach expands the
relative wave function $\psi_{\mathrm{rel}}$ 
in terms of a basis set. 
The basis functions themselves are not linearly independent,
and the determination of the eigen energies requires the
solution of a generalized eigen value problem that involves
the Hamiltonian matrix and the overlap matrix.
Just as 
with
other basis set expansion techniques, the SV approach results in
a variational upper bound to the exact eigen energies,
i.e., to the ground state energy and to the energies of excited states. 
For the interaction 
and confining potentials chosen in this work,
the functional forms of the basis functions allow
for an analytical evaluation of the Hamiltonian and overlap matrix
elements. The proper fermionic symmetry of the basis functions
is ensured
through the application of a permutation operator ${\cal{A}}$. 
For the $(3,3)$ system, e.g., ${\cal{A}}$ consists of 36
permutations (6 permutations each 
are required to anti-symmetrize the three
spin-up and the three spin-down fermions).

While the functional forms of the basis functions are relatively simple,
they are sufficiently flexible to describe short-range correlations 
that develop on a length scale of the order of 
the range $r_0$ and long-range correlations that develop
on a length scale of the order of the oscillator length
$a_{\mathrm{ho}}$~\cite{stec07c}.
This is achieved through the use of a 
comparatively
large number of variational
parameters that are optimized semi-stochastically for each basis function.
In this work, we employ basis functions that are
 characterized by $N(N-1)/2$ to $N(N-1)/2+3(N-1)$ parameters
[see Eq.~(\ref{eq_cgbasis}) 
of Appendix~\ref{sec_appendixa} for an explicit
expression for the basis functions with $L^{\Pi}=0^+$ 
symmetry and Eq.~(6.27) of Ref.~\cite{cgbook},
or Eqs.~(36) and (37) of Ref.~\cite{dail10},
for an explicit expression of the basis functions
with arbitrary $L$
employed in this work].
Generally speaking, the treatment of states with $L^{\Pi}=0^+$
is numerically less challenging than that of states with
other
symmetries. As the range $r_0$ decreases or $N$
increases, the numerical complexity of the calculation
increases. 
Also, for a given $r_0$ and $(N_1,N_2)$ combination, the numerical 
complexity increases with increasing angular momentum.
The largest calculation reported in Sec.~\ref{sec_results}
uses $N_b=3000$, where $N_b$ is the number of fully
anti-symmetrized basis functions. 
In many cases, however, 
the optimization
procedure of the variational parameters 
is more important than the size of the basis set itself. In 
our implementation, e.g., a notable 
fraction of the computational efforts
is directed at optimizing the basis functions for a given
$N_b$ as opposed to
increasing $N_b$.
The motivation for keeping the basis set relatively small
is two-fold. First, the use of highly
optimized basis functions mitigates essentially all 
problems that would otherwise arise from the linear dependence
of the basis functions~\cite{stec07c,cgbook}.
Second, the computational time required
to calculate structural properties
increases with increasing $N_b$.

To calculate structural properties, we follow two 
different approaches.
Where possible, we determine the 
matrix elements for a given operator $A$ analytically,
and determine the
quantity $\langle \psi_{\mathrm{tot}} | A | \psi_{\mathrm{tot}} \rangle / 
\langle \psi_{\mathrm{tot}} | \psi_{\mathrm{tot}} \rangle$ by simply adding
all matrix elements weighted by the appropriate 
expansion coefficients.
This approach scales quadratically with $N_b$ and is,
in most cases, more efficient than our second approch, 
a variational Monte Carlo calculation~\cite{montecarlo}
that uses the wave function optimized by the stochastic
variational approach.
In particular, we
calculate structural observables by performing a
Metropolis walk that 
samples the
probability
distribution
$|\psi_{\mathrm{tot}}|^2 / \langle \psi_{\mathrm{tot}} | \psi_{\mathrm{tot}} \rangle$.
The expectation value of $A$ is then determined by averaging
over many possible realizations of the system. At each step,
the density $|\psi_{\mathrm{tot}}|^2$ needs to be calculated,
resulting in a scaling of the computational effort
with $N_b N_{\mathrm{sample}}$, where the number of Metropolis steps
$N_{\mathrm{sample}}$ is generally much larger than $N_b$.
Appendix~\ref{sec_appendixb} details the Monte Carlo sampling scheme for 
a number of observables
that quantify the correlations of the system.
A distinct advantage of the Metropolis sampling approach is that it allows
for the evaluation of ``conditional observables''
such as the quantity $\bar{\rho}_{\mathrm{red}}(\vec{R} \, ',\vec{R})$,
defined below Eq.~(\ref{eq_reduceddensitymatrix}),
for which analytical expressions
of the matrix elements are not available.

\subsection{Density matrices, occupation numbers and
momentum distribution}
\label{sec_densitymatrix}
To quantify the correlations of trapped few-fermion systems,
we consider the radial and pair distribution functions
as well as the one- and two-body density 
matrices~\cite{lowd55,penr56,yang62,leggettbook}.
The  density matrices not
only lead to a practical route to determine the 
momentum distributions associated with the spin-up 
and spin-down atoms,
but also serve
to quantify the non-local correlations of the
system. For example, for trapped single-species Bose gases,
an eigen value of the one-body
density matrix of the order of 1 
signals a large
condensate fraction~\cite{penr56,leggettbook,dubo01,footnotenorm}.
The situation is different for two-component 
fermions~\cite{yang62,leggettbook}.
Because of the anti-symmetric many-body wave function,
none of the 
natural orbitals associated with the one-body density matrix can be
occupied macroscopically. 
In fermionic systems, an appreciable condensate fraction only
arises
if pairs are being formed~\cite{yang62,leggettbook}. 
To quantify
the correlations associated 
with the formation of pairs, one
needs to analyze the two-body density matrix.
In the following, we first introduce local structural
observables
and then non-local observables such as the
one-body density matrix
and
the two-body density matrix.
The analysis and
discussions presented 
in this paper 
are partially motivated by analogous studies of small
bosonic $^4$He and fermionic $^3$He droplets~\cite{lewa88}.
While these systems are significantly more
dense than the atomic gases considered
here, their characterization is based on the same theoretical
framework.

Specifically,
we calculate the radial density $P_{1}(\vec{r})$
for the spin-up 
atoms, where $\vec{r}$ denotes the position vector of the spin-up
atoms from the center of the trap.
The normalization is chosen
such that
\begin{eqnarray}
\label{eq_normradial}
\int P_{1}(\vec{r})  d^3\vec{r} = 1.
\end{eqnarray}
Often times, it is more convenient to record
the one-dimensional spherically symmetric component $P_{1,\mathrm{sp}}(r)$,
\begin{eqnarray}
\label{eq_radialspherical}
P_{1,\mathrm{sp}}(r)= \int P_{1}(\vec{r} \, ') \frac{\delta(r-r')}{4 \pi r'^2} 
d^3 \vec{r} \, ',
\end{eqnarray}
instead.
For $L=1$ states, e.g., the radial density $P_{1}(\vec{r})$
is not spherically symmetric, and $P_{1}(\vec{r})$
and $P_{1,\mathrm{sp}}(r)$ are different.
For spin-imbalanced systems, i.e., for systems with 
$N_1-N_2>0$, 
the radial densities for the spin-up and spin-down atoms
are different.
In this case, we also report $P_{2,\mathrm{sp}}(r)$ 
for the spin-down atoms, which 
is defined analogously to $P_{1,\mathrm{sp}}(r)$.
Similarly, we calculate the pair distribution function
$P_{12,\mathrm{sp}}(r)$ for a spin-up atom and a spin-down atom.
The normalization is the same as that for the radial
densities, i.e., Eq.~(\ref{eq_normradial}) applies if
$P_{1}(\vec{r})$ is replaced by 
$P_{12}(\vec{r})$.

In the following, we assume that the total wave function $\psi_{\mathrm{tot}}$
is
normalized to 1.
The one-body density matrix $\rho_{1}(\vec{r} \, ',\vec{r})$
for the spin-up atoms is then defined through
\begin{eqnarray}
\label{eq_onebody}
\rho_{1}(\vec{r} \, ',\vec{r})= 
\int \cdots \int 
\psi_{\mathrm{tot}}^*(\vec{r} \, ',\vec{r}_2,\cdots,\vec{r}_N)
\nonumber \\
\times
\psi_{\mathrm{tot}}(\vec{r},\vec{r}_2,\cdots,\vec{r}_N)
d^3\vec{r}_2 \cdots d^3\vec{r}_N.
\end{eqnarray}
It can be easily checked that the ``diagonal element''
$\rho_{1}(\vec{r},\vec{r})$ coincides
with the radial density $P_{1}(\vec{r})$.
The natural orbitals $\chi_{i}(\vec{r})$ can be defined
as those functions that diagonalize the one-body density 
matrix~\cite{leggettbook},
\begin{eqnarray}
\label{eq_nat}
\rho_{1}(\vec{r} \, ',\vec{r}) = 
\sum_{i} n_{i} \chi^*_{i}(\vec{r} \, ') \chi_{i}(\vec{r}),
\end{eqnarray}
where 
\begin{eqnarray}
\label{eq_nat_norm}
\int \chi^*_{i}(\vec{r}) \chi_{j}(\vec{r}) d^3\vec{r} = 
\delta_{ij}.
\end{eqnarray}
In Eq.~(\ref{eq_nat}), the $n_i$ denote the occupation numbers,
$\sum_i n_i=1$, and the subscript ``$i$'' labels the natural 
orbitals~\cite{footnoteadded}.

In practice, 
it is in general impossible to record the 
six-dimensional one-body density
matrix $\rho_{1}(\vec{r} \, ',\vec{r})$.
Thus, we define the projections
$\rho_{lm}(r',r)$,
\begin{eqnarray}
\label{eq_proj}
\rho_{lm}(r',r)=
\frac{1}{4 \pi} \nonumber \\
\times \int \int 
Y_{lm}^*(\theta',\varphi') 
\rho_{1}(\vec{r} \, ',\vec{r}) 
Y_{lm}(\theta,\varphi) 
d^2\Omega_{r'}d^2\Omega_r,
\end{eqnarray}
where $d^2 \Omega_r=\sin \theta d \theta d \varphi$.
To determine the occupation numbers and natural orbitals,
we write $\chi_i(\vec{r})=\chi_{qlm}(\vec{r})=R_{qlm}(r) Y_{lm}(\Omega_r)$
and determine the radial parts $R_{qlm}(r)$ and the occupation numbers
$n_{qlm}$ 
by diagonalizing the scaled projected density matrizes
$4 \pi \rho_{lm}(r',r)$ for each $lm$.
For a given $lm$,
$q=0$ labels the natural orbital with the largest occupation,
$q=1$ the natural orbital with the second largest
occupation, and so on.

The momentum distribution
$n_{1}(\vec{k})$ of the spin-up atoms
can be defined in terms of the 
one-body density matrix $\rho_{1}(\vec{r} \, ',\vec{r})$~\cite{leggettbook},
\begin{eqnarray}
\label{eq_momentumdist}
n_{1}(\vec{k}) = \nonumber
\\
\frac{1}{(2 \pi)^{3}} 
\int \rho_{1}(\vec{r} \, ',\vec{r}) \exp[ -i \vec{k}^T \, (\vec{r}-\vec{r} \, ')]
d^3 \vec{r} \, ' d^3 \vec{r}.
\end{eqnarray}
Using the definition of the natural orbitals $\chi_i(\vec{r})$
from Eq.~(\ref{eq_nat}),
it is shown readily that Eq.~(\ref{eq_momentumdist})
is equivalent
to
\begin{eqnarray}
\label{eq_momentum}
n_{1}(\vec{k})= \sum_i n_i |\tilde{\chi}_i(\vec{k})|^2,
\end{eqnarray}
where $\tilde{\chi}_{i}(\vec{k})$ denotes
the Fourier transform of
$\chi_{i}(\vec{r})$,
\begin{eqnarray}
\label{eq_ft}
\tilde{\chi}_i(\vec{k}) = \frac{1}{(2 \pi)^{3/2}} \int
\exp(-i \vec{k}^T \, \vec{r}) \chi_i(\vec{r}) d^3 \vec{r}.
\end{eqnarray}
As in the case of the radial density, it is convenient
to define the spherical component $n_{1,\mathrm{sp}}(k)$ of $n_{1}(\vec{k})$
through
\begin{eqnarray}
\label{eq_momentumspherical}
n_{1,\mathrm{sp}}(k) = 
\int n_{1}(\vec{k} \, ') \frac{\delta(k - k')}{4 \pi k'^2} d^3 \vec{k} \, '.
\end{eqnarray}
Appendix~\ref{sec_appendixa} 
determines analytical expressions for the matrix elements
for 
$\rho_1(\vec{r} \, ',\vec{r})$, $\rho_{lm}(\vec{r} \, ',\vec{r})$ and
$n_{1,\mathrm{sp}}(k)$
for the
basis functions that we use to
describe states with $L^{\Pi}=0^+$ symmetry.

In addition to the one-body density matrix, we consider the two-body density
matrix $\rho_{12}(\vec{r}_{\uparrow} \, ',
\vec{r}_{\downarrow} \, ',\vec{r}_{\uparrow},\vec{r}_{\downarrow})$,
\begin{eqnarray}
\rho_{12}
(\vec{r}_{\uparrow} \, ',
\vec{r}_{\downarrow} \, ',\vec{r}_{\uparrow},\vec{r}_{\downarrow})=
 \int 
\psi^*_{\mathrm{tot}}
(\vec{r}_{\uparrow} \, ',\vec{r}_2,\cdots,\vec{r}_{\downarrow} \, ',\vec{r}_{N_1+2},\cdots,\vec{r}_N)
\nonumber \\
\times
\psi_{\mathrm{tot}}
(\vec{r}_{\uparrow},\vec{r}_2,\cdots,\vec{r}_{\downarrow},\vec{r}_{N_1+2},\cdots,\vec{r}_N) 
\nonumber \\
d^3\vec{r}_2 \cdots d^3 \vec{r}_{N_1}
d^3 \vec{r}_{N_1+2} \cdots d^3 \vec{r}_{N},
\end{eqnarray}
which is obtained by integrating 
over all coordinates but the position vectors
of one of the spin-up
fermions and one of the spin-down fermions.
The two-body density matrix
quantifies the non-local correlations between 
a spin-up atom
and a spin-down atom and thus contains information about
the formation of pairs~\cite{yang62,leggettbook}.
To reduce the dimensionality of $\rho_{12}(\vec{r}_{\uparrow} \, ',
\vec{r}_{\downarrow} \, ',\vec{r}_{\uparrow},\vec{r}_{\downarrow})$,
we introduce the relative coordinate vector
$\vec{r}=\vec{r}_{\uparrow}-\vec{r}_{\downarrow}$
and the center-of-mass vector $\vec{R}=(\vec{r}_{\uparrow}+\vec{r}_{\downarrow})/2$
(and analogously for the primed coordinates), and rewrite
the two-body density matrix
in terms of these new coordinate vectors,
i.e., we transform to a new set of coordinates.
We then define the reduced two-body density matrix
$\rho_{\mathrm{red}}(\vec{R} \, ',\vec{R})$ 
through
\begin{eqnarray}
\label{eq_reduceddensitymatrix}
\rho_{\mathrm{red}}(\vec{R} \, ',\vec{R})= \nonumber \\
\int \rho_{12}
\left(\vec{R} \, '+\frac{\vec{r}}{2},\vec{R} \, '-\frac{\vec{r}}{2},
\vec{R}+\frac{\vec{r}}{2},\vec{R}-\frac{\vec{r}}{2} \right)
d^3 \vec{r}.
\end{eqnarray}
The quantity $\rho_{\mathrm{red}}(\vec{R} \, ',\vec{R})$ measures the non-local
correlations between
spin-up---spin-down
pairs that are characterized by the same relative 
distance vector
$\vec{r}$.
However, as defined the reduced two-body density matrix
$\rho_{\mathrm{red}}(\vec{R} \, ',\vec{R})$ does not distinguish
between ``small'' and ``large'' pairs.
For sufficiently small $a_s$ ($a_s>0$), we expect that the system consists of
$N_2$ point-like pairs and $N_1-N_2$ unpaired atoms and that the $N_2$ pairs
form a molecular Bose gas. The $(3,2)$ system, e.g., 
can be thought of as consisting of
two pairs and one fermionic impurity
when $a_s$ is small ($a_s >0$);
in this case, the pair fraction should be determined by 
the non-local correlations of the two composite molecules
(as opposed to the non-local correlations of all $N_1N_2=6$ possible
pairs).
Thus,
we define the quantity $\bar{\rho}_{\mathrm{red}}(\vec{R} \, ',\vec{R})$,
which is obtained from
$\rho_{\mathrm{red}}(\vec{R} \, ',\vec{R})$
by only including those
$\vec{R}$-vectors 
that correspond to the
position vectors of one of the smallest
$N_2$ pairs.
In practice, we determine
$\bar{\rho}_{\mathrm{red}}(\vec{R} \, ',\vec{R})$ 
during the Metropolis walk
(see Sec.~\ref{sec_densitymatrix} and Appendix~\ref{sec_appendixb}). 
While $\rho_{\mathrm{red}}(\vec{R} \, ',\vec{R})$
is sampled at each step,
$\bar{\rho}_{\mathrm{red}}(\vec{R} \, ',\vec{R})$
is only sampled if the $\vec{R}$ under consideration
belongs to that of one of the $N_2$ smallest pairs.
We note that a related approach has been employed in the
Monte Carlo treatment of one-dimensional 
spin-imbalanced Fermi gases~\cite{casu08}.

Just as with the one-body density matrix $\rho_1(\vec{r} \, ',\vec{r})$,
the reduced two-body density matrix $\rho_{\mathrm{red}}(\vec{R} \, ',\vec{R})$
can
be decomposed into natural orbitals $\chi_{qlm}(\vec{R})$.
We refer to the corresponding occupation numbers as $N_{qlm}$; the capital
$N$ is chosen to distinguish the occupation numbers associated
with $\rho_{\mathrm{red}}(\vec{R} \, ',\vec{R})$ 
from those associated with $\rho_1(\vec{r} \, ',\vec{r})$.
In analogy to the formalism outlined
above for the one-body density matrix, 
we define the projections $\rho_{lm}(R',R)$
and $\bar{\rho}_{lm}(R',R)$ of the reduced 
two-body density matrix.
While the occupation numbers $N_{qlm}$ obtained by
diagonalizing the $\rho_{lm}(R',R)$ add up,
 by construction, to 1, those obtained by diagonalizing 
the $\bar{\rho}_{lm}(R',R)$ do not. This is a direct consequence
of the ``conditional sampling approach''.
Appendices~\ref{sec_appendixb} and \ref{sec_appendixc}
provide more details about the Monte Carlo sampling and the 
behavior of ${\rho}_{lm}(R',R)$ 
and $\bar{\rho}_{lm}(R',R)$ in the $a_s \rightarrow 0^+$ limit.

In analogy to Eq.~(\ref{eq_momentumdist}), 
the reduced two-body density matrix can
be used
to obtain the momentum distribution $n_{\mathrm{red}}(\vec{K})$;
here, we use $\vec{K}$ instead of $\vec{k}$ to
distinguish the momentum vector associated with the 
position vector of a pair from that of an
atom. Similarly, we define
$n_{\mathrm{red},\mathrm{sp}}(K)$.

\section{Results}
\label{sec_results}

This section presents our results for small trapped 
two-component Fermi gases with equal masses.
We first present results for the energies of systems with
up to $N=6$ particles (see Sec.~\ref{sec_energetics})
and then discuss selected local
structural properties (see
Sec.~\ref{sec_structure}).
Lastly, Sec.~\ref{sec_structure2} discusses our
results obtained by analyzing non-local observables.

\subsection{Energetics}
\label{sec_energetics}
The energetics of the $(2,1)$ and $(2,2)$ systems have
been discussed in detail in 
the literature.
Here,
we focus on the $(3,2)$ and $(3,3)$ systems.
While the qualitative behavior of these larger systems
is similar to that of the three- and four-particle systems, 
the energy spectra of the larger systems is more complex.
The increase of the complexity can be traced back to
the increased degeneracies in the limits
that $a_s \rightarrow 0^-$ and $a_s \rightarrow 0^+$.
In the weakly-attractive regime
($a_s < 0$ and $|a_s|/a_{\mathrm{ho}} \ll 1$), 
the so-called BCS regime, the
system behaves like a weakly-attractive atomic Fermi 
gas (see, e.g., Ref.~\cite{gior08}). In the weakly-repulsive
regime
($a_s > 0$ and $a_s/a_{\mathrm{ho}} \ll 1$), 
the so-called BEC regime, the system behaves like a weakly-repulsive
molecular Bose gas with $N_1-N_2$ unpaired ``fermionic 
impurities'' (see, e.g., Ref.~\cite{gior08}).
The degeneracies of the non-interacting atomic Fermi gas and
the molecular Bose gas with fermionic impurities
can be obtained by extending the hyperspherical framework discussed 
in Ref.~\cite{dail10} for the $(2,1)$ and $(2,2)$
systems to larger systems.
Furthermore, the lifting of the degeneracies, i.e.,
the slope of each energy level for small $|a_s|$, $a_s < 0$ and $a_s>0$,
can be obtained by applying first order degenerate perturbation
theory using Fermi's pseudo-potential~\cite{stec07b,stec08,dail10}. 
While these limiting behaviors can be obtained fairly straightforwardly,
the behavior of
the energy levels in the strongly-correlated
regime,
i.e., in the regime where $|a_s|/a_{\mathrm{ho}} \gtrsim 1$,
is, in general,
non-trivial.
In the following, we highlight selected features of
the energy spectra of the $(3,2)$ and $(3,3)$
systems. 

\begin{figure}
\vspace*{+.9cm}
\includegraphics[angle=0,width=60mm]{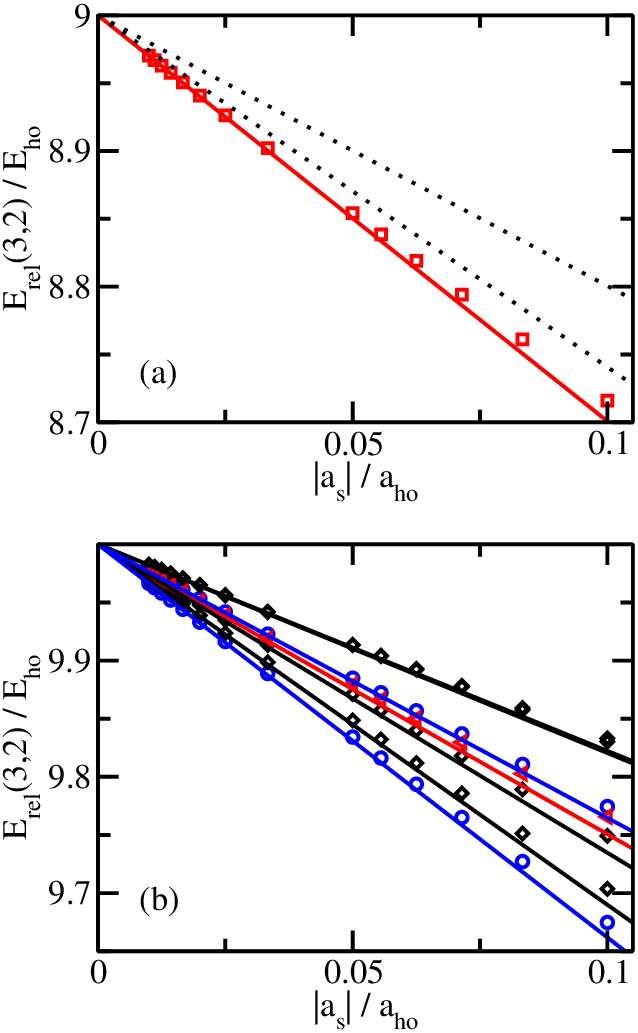}
\vspace*{0.2cm}
\caption{(Color online)
Energies $E_{\mathrm{rel}}$ of the $(3,2)$ system in the weakly-attractive
regime as a function of $|a_s|$, $a_s < 0$.
(a) Energy manifold around $E_{\mathrm{rel,ni}}=9 \hbar \omega$.
Squares show the SV energies for the state with $1^-$ symmetry
while a solid line shows the energies obtained
within first order degenerate perturbation theory.
Dotted lines show the perturbative energies for the unnatural
parity states with
$0^-$ symmetry (upper curve) and $2^-$ symmetry
(lower curve with 5-fold degeneracy).
(b) Energy manifold around $E_{\mathrm{rel,ni}}=10 \hbar \omega$.
Circles, diamonds and triangles  
show the SV energies for the states with $0^+$
symmetry (two levels with 1-fold degeneracy each),
$2^+$ symmetry (four levels with 5-fold degeneracy each; 
the upper two curves
are nearly degenerate) and $4^+$ symmetry
(one level with 9-fold degeneracy), respectively,
while solid lines show the energies obtained
within first order degenerate perturbation theory.
The unnatural parity states with $1^+$ 
and $3^+$ symmetry 
are not shown.
The SV calculations are performed for $r_0=0.05a_{\mathrm{ho}}$.
The harmonic oscillator energy $E_{\mathrm{ho}}$ is defined as 
$E_{\mathrm{ho}}=\hbar \omega$.
}\label{fig_en5_pert}
\end{figure}

\begin{figure}
\vspace*{+.9cm}
\includegraphics[angle=0,width=65mm]{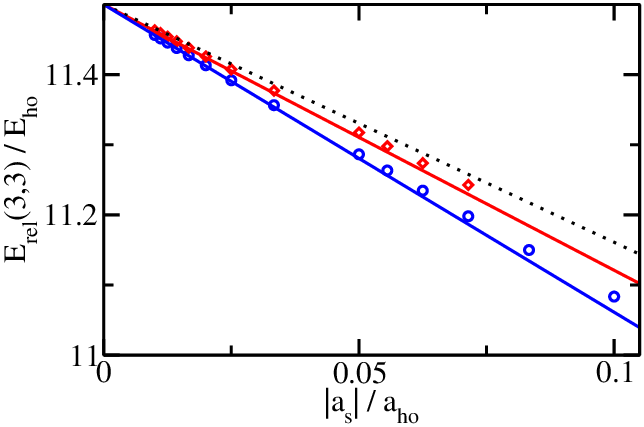}
\vspace*{0.2cm}
\caption{(Color online)
Lowest energy manifold of the $(3,3)$ system in the weakly-attractive
regime as a function of $|a_s|$, $a_s<0$.
Circles and diamonds show the 
SV energies for the natural parity states with 
$0^+$ and $2^+$ symmetry, respectively,
while solid lines show the energies calculated using first
order degenerate perturbation theory.
In addition, a dotted line shows the perturbative
energy for the unnatural parity state with $1^+$ symmetry.
The SV calculations are performed for $r_0=0.05a_{\mathrm{ho}}$.
}\label{fig_en6_pert}
\end{figure}

Figure~\ref{fig_en5_pert} shows the energies
of the $(3,2)$ system
in the weakly-attractive
regime as a function of $|a_s|$
for the first two energy manifolds around the
non-interacting energies $E_{\mathrm{rel,ni}}=9 \hbar \omega$
and $E_{\mathrm{rel,ni}}=10 \hbar \omega$.
These energy manifolds consist of a total of
9 and 
57 states, respectively
(see Table~\ref{tab_en5_pert}).
\begin{table}
\caption{Dimensionless coefficients $c^{(1)}$
that characterize the weakly-attractive 
Fermi gas for the
$(N_{1},N_{2})=(3,2)$ system.
The $c^{(1)}$ are defined through
$E^{(1)}= c^{(1)} (2 \pi)^{-1/2} \hbar \omega a_s/a_{\mathrm{ho}}$,
where $E^{(1)}$ denotes the first order perturbative energy shift,
i.e., $E_{\mathrm{rel}} \approx E_{\mathrm{rel,ni}}+E^{(1)}$.
$E_{\mathrm{rel,ni}}$ denotes the relative energy of the non-interacting
system and $g_{\mathrm{rel,ni}}$ the degeneracy, 
i.e., $g_{\mathrm{rel,ni}}=2L+1$.}
\begin{ruledtabular}
\begin{tabular}{cccc} 
$E_{\mathrm{rel,ni}}/(\hbar\omega)$ & $g_{\mathrm{rel,ni}}$ & 
$L^\pi$ & $c^{(1)}$ \\ \hline
9	& 5	& $2^-$	& $13/2$ 		\\
9	& 3	& $1^-$	& $15/2 $		\\
9	& 1	& $0^-$	& $5		$\\ \hline
10	& 9	& $4^+$	& $25/4		$\\
10	& 7	& $3^+$	& $21/4		$\\
10	& 7	& $3^+$	& $9/2		$\\
10	& 5	& $2^+$	& $7.77155	$\\
10	& 5	& $2^+$	& $6.65010	$\\
10	& 5	& $2^+$	& $9/2		$\\
10	& 5	& $2^+$	& $4.45335	$\\
10	& 3	& $1^+$	& $\frac{1}{16} (87+\sqrt{209})$ \\
10	& 3	& $1^+$	& $\frac{1}{16} (87-\sqrt{209})$ \\
10	& 3	& $1^+$	& $\frac{1}{8} (33+\sqrt{89}) $\\
10	& 3	& $1^+$	& $\frac{1}{8} (33-\sqrt{89}) $\\
10	& 1	& $0^+$	& $\frac{5}{16} (23+\sqrt{17})$ \\
10	& 1	& $0^+$	& $\frac{5}{16} (23-\sqrt{17})$ \\ 
\end{tabular}
\end{ruledtabular}
\label{tab_en5_pert}
\end{table}
For comparison, the lowest energy manifold of
the $(2,1)$
and $(2,2)$ systems
contains only 3 and 9 states, respectively, and the second lowest
energy manifold of these systems 
contains only 9 and 27 states, respectively~\cite{dail10}.
The ground state of the ($3,2)$ system has $L^{\Pi}=1^-$ symmetry and is
3-fold degenerate (the degeneracy $g_{\mathrm{rel,ni}}=3$ is due to the spherical 
symmetry and is associated with the azimuthal quantum
number $M_L$, $M_L=-L,-L+1,\cdots,L$).
The two excited states of the lowest energy manifold 
[dotted lines in Fig.~\ref{fig_en5_pert}(a)] 
correspond to unnatural parity states with $0^-$ 
and $2^-$ symmetry.
For $|a_s|/a_{\mathrm{ho}} \ll 1$, the perturbative treatment describes
the energy spectrum accurately.  
As expected,
the description worsens as $|a_s|/a_{\mathrm{ho}}$ increases.
We note that the
finite-range effects of the SV energies are smaller than the symbol
size; consequently, the deviations between 
the SV energies and the perturbative energies
are predominantly due to the approximate nature of the
perturbative treatment, which assumes zero-range interactions, and not
due to the fact that Fig.~\ref{fig_en5_pert} compares energies 
obtained for finite-range and zero-range interactions.
The perturbative treatment 
provides a qualitatively correct
picture up to $|a_s|/a_{\mathrm{ho}} \approx 0.5$
(note that Fig.~\ref{fig_en5_pert}
only covers the values $|a_s|/a_{\mathrm{ho}} \le 0.1$).
Figure~\ref{fig_en5_pert}(b) 
shows the energy levels 
corresponding to natural parity states
of the first excited state energy manifold
of the $(3,2)$ system
around $E_{\mathrm{rel,ni}}=10 \hbar \omega$.
For comparison, 
Fig.~\ref{fig_en6_pert}
exemplarily illustrates 
for the
$(3,3)$ system that the ground state of spin-balanced systems
has $0^+$ symmetry. 
Table~\ref{tab_en6_pert} summarizes the degeneracies and 
\begin{table}
\caption{Dimensionless coefficients $c^{(1)}$
for the Fermi gas with $(N_1,N_2)=(3,3)$.
See the caption of Table~\ref{tab_en5_pert} for details.}
\begin{ruledtabular}
\begin{tabular}{cccc} 
$E_{\mathrm{rel,ni}}/(\hbar\omega)$ & $g_{\mathrm{rel,ni}}$ & $L^\pi$ & $c^{(1)}$ \\ \hline
23/2	& 5	& $2^+$	& $19/2$		\\
23/2	& 3	& $1^+$	& $17/2	$	\\
23/2	& 1	& $0^+$	& $11		$\\ \hline
25/2	& 9	& $4^-$	& $33/4		$\\
25/2	& 9	& $4^-$	& $31/4		$\\
25/2	& 7	& $3^-$	& $41/4		$\\
25/2	& 7	& $3^-$	& $\frac{1}{8} (63+\sqrt{33}) $\\
25/2	& 7	& $3^-$	& $\frac{1}{8} (63-\sqrt{33}) $\\
25/2	& 5	& $2^-$	& $73/8	$	\\
25/2	& 5	& $2^-$	& $9.04636	$\\
25/2	& 5	& $2^-$	& $17/2		$\\
25/2	& 5	& $2^-$	& $8.11599	$\\
25/2	& 5	& $2^-$	& $6.71264	$\\
25/2	& 5	& $2^-$	& $11/2		$\\
25/2	& 3	& $1^-$	& $\frac{1}{16} (137+\sqrt{609}) $\\
25/2	& 3	& $1^-$	& $9.34613	$\\
25/2	& 3	& $1^-$	& $8.58664	$\\
25/2	& 3	& $1^-$	& $\frac{1}{16} (137-\sqrt{609})$ \\
25/2	& 3	& $1^-$	& $5.94223	$\\
25/2	& 1	& $0^-$	& $7		$\\
25/2	& 1	& $0^-$	& $25/4		$\\ 
\end{tabular}
\end{ruledtabular}
\label{tab_en6_pert}
\end{table}
perturbative energy shifts 
for the two lowest energy manifolds of the $(3,3)$ system.

\begin{figure}
\vspace*{+.9cm}
\includegraphics[angle=0,width=65mm]{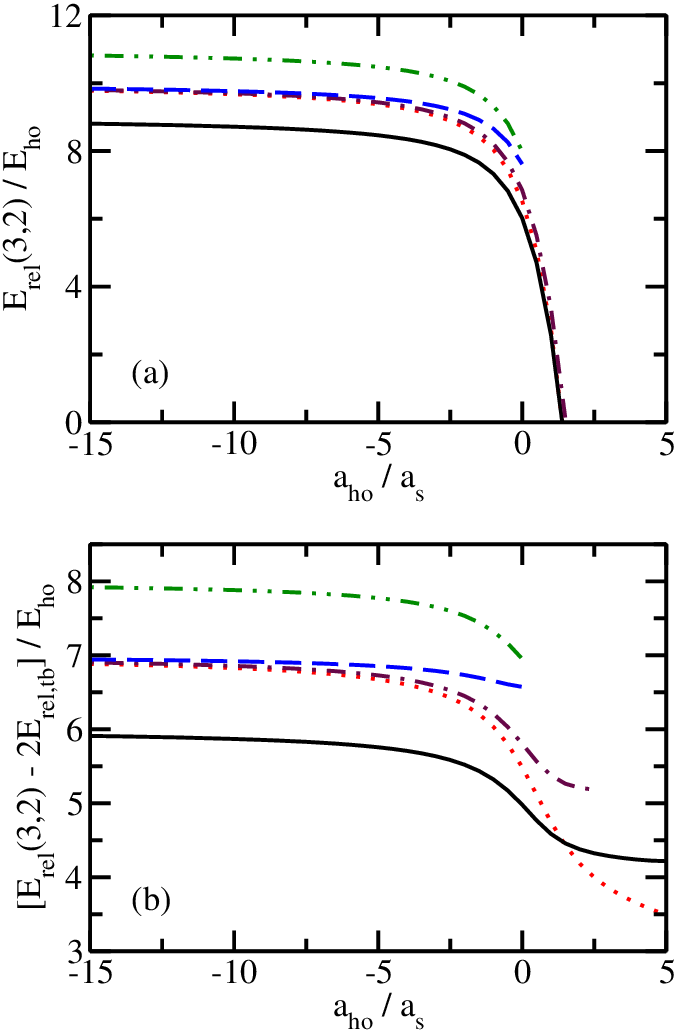}
\vspace*{0.2cm}
\caption{(Color online)
SV energies for the natural parity states
of the $(3,2)$ system with $r_0=0.05a_{\mathrm{ho}}$
as a function of $a_s^{-1}$ in the crossover regime.
Panel (a) shows the ``bare energy'' $E_{\mathrm{rel}}$ while
panel (b) shows the scaled energy $E_{\mathrm{rel}}-2E_{\mathrm{rel,tb}}$.
Dotted, solid, dash-dotted, dash-dot-dotted and dashed
lines correspond to the lowest state with $L^{\Pi}=0^+$,
$1^-$, $2^+$, $3^-$ and $4^+$ symmetry, respectively.
The $L=2-4$ curves do not extend all the way to $a_{\mathrm{ho}}/a_s=10$
since the convergence of the energies on 
the positive scattering length
side becomes more challenging as $L$ increases.
}\label{fig_en5_crossover}
\end{figure}

Figure~\ref{fig_en5_crossover} shows selected
energy levels for natural parity states
of the $(3,2)$ system as a function of $a_s^{-1}$ throughout
the crossover. 
Dotted, solid, dash-dotted, dash-dot-dotted 
and dashed lines
show the lowest energy level of the $L=0$ to 4 states
with natural parity.
Figure~\ref{fig_en5_crossover}(a) 
shows that the $L=1$ state has the lowest energy 
when $a_s$ is negative [see also Fig.~\ref{fig_en5_pert}(a)].
However, when $a_s$ is small and positive, the $L=0$ state
has lower energy. 
This can be most clearly seen in Fig.~\ref{fig_en5_crossover}(b),
which shows the scaled energy $E_{\mathrm{rel}}-2 E_{\mathrm{rel,tb}}$,
where $E_{\mathrm{rel,tb}}$ denotes the relative ground state energy
of two  trapped atoms that interact through the same two-body potential
as the corresponding five-particle system.
The subtraction of the energy of two dimers is motivated by
the fact that the fermionic system behaves like a system 
that consists of
$N_2$ diatomic molecular bosons and $N_1-N_2$ 
fermions~\cite{astr04c,stec07b,stec08,kest07}.
By subtracting the ``internal'' two-body binding energy $E_{\mathrm{rel,tb}}$, 
the energy crossover curves are mapped to a smaller
energy interval which more clearly reveals the key physics.
For example, a significant fraction of the 
finite-range effects on the positive
scattering length side arises due to the formation of pairs
and
is 
removed by
subtracting the binding energy
of $N_2$ dimers.
Figure~\ref{fig_en5_crossover}(b) shows that
the crossing between the $L^{\Pi}=1^-$ and $0^+$ curves
occurs
at $a_{\mathrm{ho}}/a_s\approx 1.5$ for the $(3,2)$ system. 
This is slightly larger than the value
at which the crossing occurs for the $(2,1)$ system,
i.e., 
$a_{\mathrm{ho}}/a_s \approx 1$~\cite{kest07,stet07,stec08}.

\begin{figure}
\vspace*{+.9cm}
\includegraphics[angle=0,width=65mm]{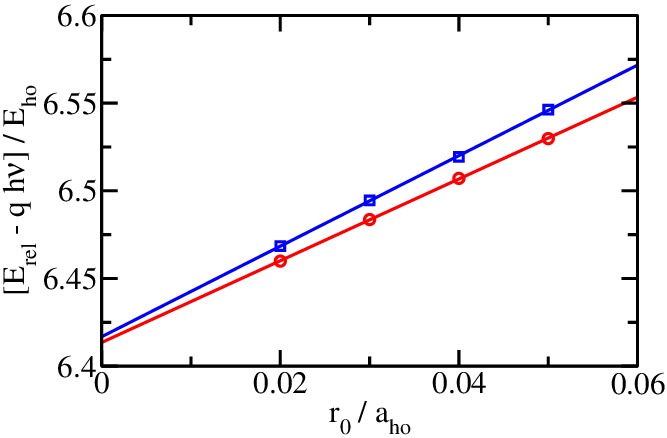}
\vspace*{0.2cm}
\caption{(Color online)
Energetics of the $(3,2)$ system with $L^{\Pi}=0^+$ at unitarity as a function
of $r_0$.
Circles 
show the SV energy $E_{\mathrm{rel}}$ ($q=0$ on the $y$-axis label)
for the lowest state with $L^{\Pi}=0^+$ symmetry,
while squares show the shifted SV energy $E_{\mathrm{rel}}-2 \hbar \omega$
($q=1$ on the $y$-axis label)
of the second excited state.
Solid lines show a linear fit to the SV energies.
The intercepts $E_{\mathrm{rel}}(r_0=0)$ and slopes
are
$6.4135(7) \hbar \omega$
and $2.33(2) \hbar \omega /r_0$
for the ground state,
and  
$6.417(1) \hbar \omega$
and $2.58(3) \hbar \omega /r_0$
for the second excited state, respectively.
The numbers in brackets reflect the uncertainty arising from the fit
and neglect the basis set extrapolation error of the SV energies.
}\label{fig_n5_excited}
\end{figure}

\begin{figure}
\vspace*{+.9cm}
\includegraphics[angle=0,width=65mm]{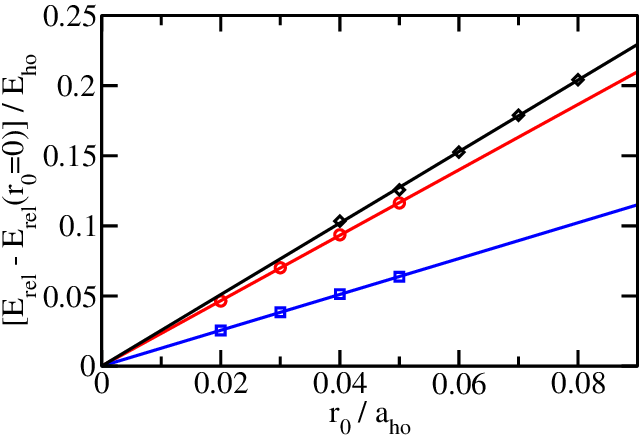}
\vspace*{0.2cm}
\caption{(Color online)
Shifted energies 
$E_{\mathrm{rel}}-E_{\mathrm{rel}}(r_0=0)$
at unitarity  as a function of 
$r_0$.
Circles and squares show the shifted energy of the lowest
state of the $(3,2)$ 
system with $L^{\Pi}=0^+$ and $1^-$ symmetry, respectively,
while 
diamonds show the shifted energy of the lowest
state of the $(3,3)$ 
system with $L^{\Pi}=0^+$.
Solid lines show linear fits to the SV energies.
}\label{fig_range}
\end{figure}

We now discuss 
the infinite scattering length regime,
which has received considerable
attention for several reasons. On the one hand, this is
the regime where the system is most
strongly
correlated and where no small parameter
exists around which to expand.
On the other hand, the very same aspect that
leads to the strong correlations, 
namely the infinitely large $s$-wave scattering
length, also leads to a scale invariance of the 
system~\cite{cast04,wern06}.
In the zero-range limit, the unitary system
is characterized by the same number 
of length scales as the 
non-interacting system, which 
can be shown to imply the separability of the 
wave function into a hyperradial part and a hyperangular
part~\cite{cast04,wern06}.
This separability has a number of consequences. One of these 
is the existence of ladders of energy levels that
are separated by $2 \hbar \omega$~\cite{wern06,blum07}.
Figure~\ref{fig_n5_excited} 
exemplarily illustrates 
for the $(3,2)$ system
with $L^{\Pi}=0^+$
symmetry
how this $2 \hbar \omega$
spacing changes as a function of the range $r_0$ of the two-body
interaction potential. 
Circles show the
ground state energy while squares show the energy of the second excited 
state, with $2 \hbar \omega$ subtracted, for various $r_0$.
Figure~\ref{fig_n5_excited} shows that
the finite-range energies approach the zero-range limit linearly  from above.
The two-paramater fits, shown by solid lines, nearly coincide 
at $r_0=0$, numerically confirming the expected
$2 \hbar \omega$ spacing with better than $0.1\%$ accuracy. 
Assuming that a numerically exact treatment gives 
$E_{\mathrm{rel,gr}}-E_{\mathrm{rel,exc}}=2\hbar\omega$ for $r_0=0$, 
Fig.~\ref{fig_n5_excited}
can be used to
assess the accuracy of the SV energies and 
the extrapolation scheme.
Figure~\ref{fig_range} shows additional examples for the 
range dependence of the few-body energies at unitarity. 

Table~\ref{tab_angmom}
summarizes the extrapolated zero-range energies for $N=4-6$.
\begin{table}       
\caption{Natural parity zero-range energies
$E_{\mathrm{rel}}(N_1,N_2)$, in units    
of $\hbar \omega$, for the two-component 
equal-mass Fermi gas at unitarity.   
The energies are obtained by solving a  
transcendental equation
for $N=3$~\protect\cite{wern06a}.
For $N=4,5$ and 6, the energies
are obtained by analyzing the            
SV energies for finite $r_0$:
The first entry in the third through fifth 
column is obtained by extrapolating the
lowest SV energy for each $r_0$ to the $r_0 \rightarrow 0$ limit
(the results for $N=4$ are taken from Ref.~\protect\cite{dail10}).
The second entry in the third through fifth 
column is obtained by first extrapolating
the SV energies to the $N_b \rightarrow \infty$ limit
for each $r_0$ and by then extrapolating the resulting energies
to the $r_0 \rightarrow 0$ limit.                                           
} 
\begin{ruledtabular}
\begin{tabular}{l|c|ccc}  
$L^{\Pi}$ & $E_{\mathrm{rel}}(2,1)$ & $E_{\mathrm{rel}}(2,2)$ & $E_{\mathrm{rel}}(3,2)$ & $E_{\mathrm{rel}}(3,3)$ \\
\hline        
$0^+$ & 3.166 & 3.509/3.509 & 6.413/6.395 & 6.858/6.842 \\  
$1^-$ & 2.773 & 5.598/5.596 & 5.958/5.955 & 8.742/8.682 \\  
$2^+$ & 4.105 & 4.418/4.418 & 6.775/6.774 & 7.855/7.829 \\
$3^-$ & 4.959 & 6.176/6.174 & 7.906/7.898 & 8.279/8.269 \\  
$4^+$ & 6.019 & 6.485/6.484 & 7.603/7.601 & 9.569/9.534 \\ 
$5^-$ & 6.992 & 8.245/8.243 & 8.955/8.945 & 10.43/10.40 \\ 
$6^+$ & 8.004 & 8.496/8.496 & 9.657/9.653 & 10.36/10.32  
\end{tabular}                
\end{ruledtabular}            
\label{tab_angmom} 
\end{table}
In analyzing our finite range
SV energies, we pursued two approaches: The first approach determines
the
$r_0 \rightarrow 0$ energies by
fitting a linear curve to the lowest SV energies 
for between 2 and 5 different $r_0$ 
(the results are given by
the first entry in the third through fifth 
column in Table~\ref{tab_angmom}). 
The second approach first extrapolates the SV energies for each $r_0$
to the infinite basis set 
limit, i.e., to the $N_b \rightarrow \infty$ limit, 
and then determines
the $r_0 \rightarrow 0$ energies by fitting the extrapolated SV energies
(the results are given
by the second entry in the third through fifth 
column in Table~\ref{tab_angmom}).
As can be seen, the energies obtained by the second approach
lie, as expected, below the energies obtained by the first approach.
The second entry in the third through fifth column is
our best estimate for the zero-range energy.
The errorbars depend on both extrapolations conducted and
are not entirely straightforward to determine reliably.
For $N=4$ and $L>0$, we estimate the uncertainties to be 
the larger of $0.005 \hbar \omega$ and the absolute
value of the difference of the two entries in column three
(for $N=4$ and $L=0$, the uncertainty is $0.001\hbar \omega$).
For $N=5$ ($N=6$), we estimate the uncertainties to be 
the larger of $0.01 \hbar \omega$ ($0.02 \hbar \omega$)
and the absolute
value of the difference of the two entries in column 
four (five).

While the range
dependence 
at unitarity
varies notably with the symmetry of the
system, the energy increases with increasing $r_0$ for all
systems considered in Table~\ref{tab_angmom}.
In particular, we find that the slopes vary
between about $0.08 \hbar \omega/r_0$ and 
about $2.50 \hbar \omega/r_0$. 
While the range dependence does, of course,
depend on the shape of the two-body potential, we believe that the
range dependence 
for other short-range model potentials is similar to that found here 
for the Gaussian interaction potential. A more detailed discussion of
the dependence of the energies on the range of the two-body potential
or the effective range, which characterizes the leading order energy 
dependence of the two-body $s$-wave phase shift, can be found in 
Refs.~\cite{efim93,wern08,wern10}.

\begin{figure}
\vspace*{+.9cm}
\includegraphics[angle=0,width=65mm]{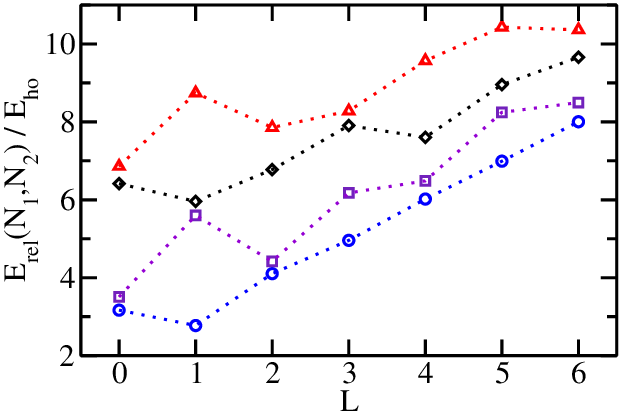}
\vspace*{0.2cm}
\caption{(Color online)
Circles, squares, diamonds and triangles 
show the extrapolated
zero-range energy $E_{\mathrm{rel}}(N_1,N_2)$ at unitarity as a function of $L$
for the $(2,1)$, $(2,2)$, $(3,2)$, and $(3,3)$ systems, respectively.
For each $L$, the energy of the energetically lowest lying natural parity
state is shown. 
Dotted lines are shown to guide the eye.
The energies are listed in Table~\ref{tab_angmom}.
}\label{fig_zrenergy}
\end{figure}

Figure~\ref{fig_zrenergy} shows the energies of
Table~\ref{tab_angmom} graphically.
While we were able to interpret the energies of the $(2,1)$ and $(2,2)$
systems within a simple model (see Ref.~\cite{dail10}),
we did not find simple analytical 
expressions that would predict the  energies of the $(3,2)$ and $(3,3)$
systems at unitarity with a few percent accuracy.
The energies summarized in Table~\ref{tab_angmom}
are, to the best of our knowledge, the most extensive
and precise
esimates of the zero-range energies for systems with $N=5$ and 6,
and can be used to assess the accuracy of other numerical approaches.
For example, the fixed-node Monte Carlo energies presented in 
Refs.~\cite{blum07,stec08} 
for a square well potential with range $0.01a_{\mathrm{ho}}$
are between 0.1\% and 4\% higher than the zero-range energies
reported in the first entry of columns three 
to five of Table~\ref{tab_angmom}. 
We estimate that roughly up to 1\% of the deviations can be
attributed to finite-range effects. The remaining discrepancy
suggests that 
the nodal surfaces employed in the
fixed-node Monte Carlo calculations are not
perfect.

\subsection{Local structural properties}
\label{sec_structure}

This section characterizes local structural properties of
small two-component Fermi gases.
As discussed in Sec.~\ref{sec_energetics}, the ground state
of spin-imbalanced systems with $N_1-N_2=1$ has $1^-$ symmetry
in the weakly-attractive regime and $0^+$ symmetry in the
weakly-repulsive regime, while the ground state of spin-balanced systems
has $0^+$ symmetry throughout the entire crossover.
Motivated by this observation, this section focuses on
the energetically lowest lying 
states with $L^{\Pi}=0^+$ and $1^-$ symmetry. 

\begin{figure}
\vspace*{+.9cm}
\includegraphics[angle=0,width=60mm]{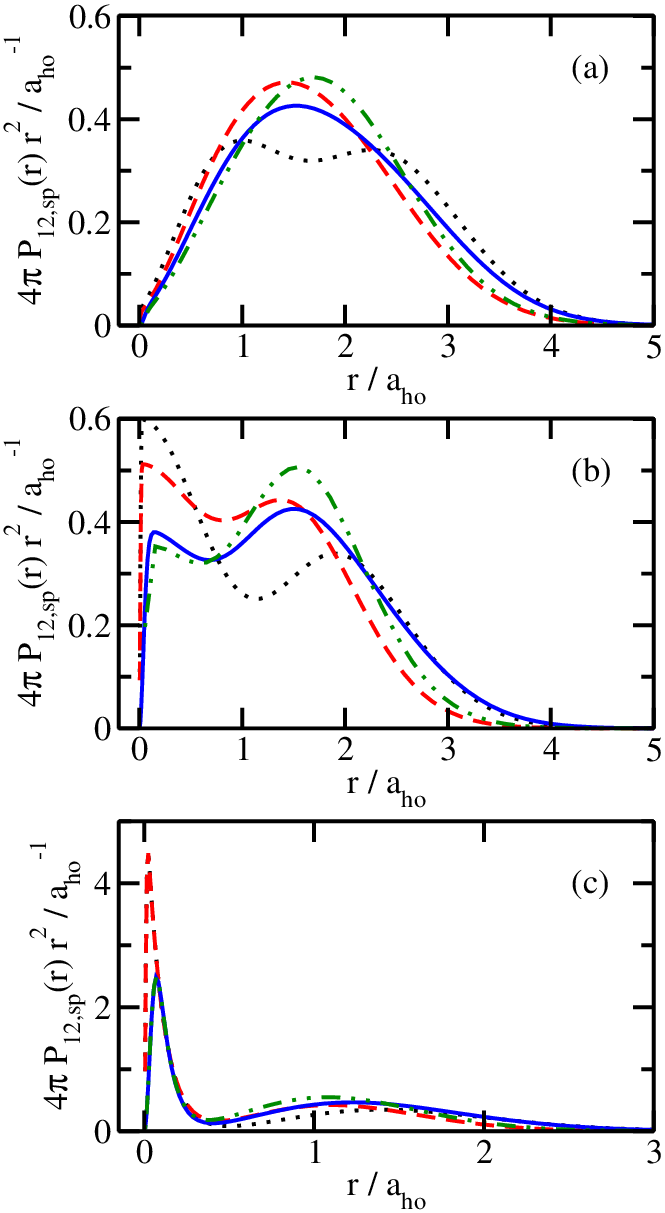}
\vspace*{0.2cm}
\caption{(Color online)
Scaled pair distribution function $4 \pi \, P_{12,\mathrm{sp}}(r)r^2$
for the lowest $L^{\Pi}=0^+$ state of
the 
$(2,1)$ system (dotted lines),
the $(2,2)$ system (dashed lines),
the $(3,2)$ system (solid lines) and
the $(3,3)$ system (dash-dot-dotted lines)
for (a) $a_{\mathrm{ho}}/a_s=-5$,
(b) $a_{\mathrm{ho}}/a_s=0$ and
(c) $a_{\mathrm{ho}}/a_s=5$.
The calculations for the $(2,1)$ and $(2,2)$
systems are performed using $r_0=0.01a_{\mathrm{ho}}$
while those for the $(3,2)$ and $(3,3)$
systems are performed using $r_0=0.05a_{\mathrm{ho}}$.
The pair distribution functions for the $(2,1)$ and $(2,2)$ 
systems at unitarity agree with those
presented in Ref.~\protect\cite{stec08}.
}\label{fig_pairdist_l0}
\end{figure}

Figure~\ref{fig_pairdist_l0} shows
the pair distribution function $P_{12,\mathrm{sp}}(r)$
for the 
$(2,1)$ system (dotted lines), 
the $(2,2)$ system (dashed lines), 
the $(3,2)$ system (solid lines),
and the $(3,3)$ system (dash-dot-dotted lines)
with $0^+$ symmetry.
Figure~\ref{fig_pairdist_l0}(a), (b) and (c)
show the pair distribution functions for
$a_{\mathrm{ho}}/a_s=-5$, $0$ and 5, respectively. 
While the overall behavior of the pair distribution functions
for different $N$ but fixed $a_s/a_{\mathrm{ho}}$ is similar, small 
differences
exist.
For example, for all scattering lengths,
the scaled pair distribution
functions of the spin-balanced 
$(2,2)$ and $(3,3)$ systems
take on vanishingly small values at smaller $r$  than those of
the spin-imbalanced $(2,1)$ and $(3,2)$ systems.
This behavior is reversed for the $L^{\Pi}=1^-$ states (see
Fig.~\ref{fig_pairdist_l1}).
The scaled pair distribution functions $P_{12,\mathrm{sp}}(r)r^2$
for $a_{\mathrm{ho}}/a_s=-5$ [Fig.~\ref{fig_pairdist_l0}(a)]
have a small but non-vanishing 
amplitude for $r$ values of the order of $r_0$,
reflecting the weakly-attractive nature of the
two-body interactions. 
For $a_{\mathrm{ho}}/a_s=0$ and $5$, the scaled pair
distribution functions
$P_{12,\mathrm{sp}}(r)r^2$
are characterized by two peaks. As discussed in detail in 
Ref.~\cite{stec08} for the
$(2,1)$ and $(2,2)$ systems, the two-peak structure arises
due to the formation of pairs. 
While both peaks are broad at unitarity [Fig.~\ref{fig_pairdist_l0}(b)],
the peak at smaller $r$ becomes notably more pronounced as
the scattering length becomes positive
[Fig.~\ref{fig_pairdist_l0}(c)]. This can be understood intuitively
by realizing that the size of the pairs is,
for sufficiently small $a_s$ ($a_s$ positive), set by $a_s$, thereby
giving rise to the pronounced peak of $P_{12,\mathrm{sp}}(r)r^2$
around $r \approx a_s$.
The fact that the scaled pair distribution functions
go to 0 as $r \rightarrow 0$ is due to the use of
finite-range interaction potentials. If we had used zero-range 
interactions,
the amplitude of $P_{12,\mathrm{sp}}(r)r^2$ would be finite
at $r=0$.

\begin{figure}
\vspace*{+.9cm}
\includegraphics[angle=0,width=60mm]{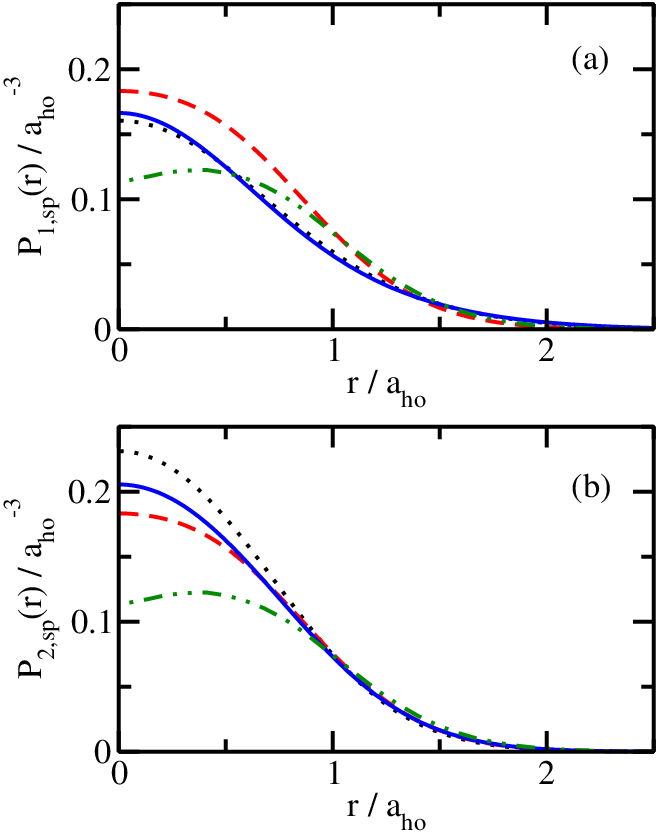}
\vspace*{0.2cm}
\caption{(Color online)
Panels (a) and (b) show
the radial densities $P_{1,\mathrm{sp}}(r)$
and $P_{2,\mathrm{sp}}(r)$, respectively,
for the lowest $L^{\Pi}=0^+$ state 
of the
$(2,1)$ system (dotted lines),
the $(2,2)$ system (dashed lines),
the $(3,2)$ system (solid lines) and
the $(3,3)$ system (dash-dot-dotted lines)
at unitarity.
The calculations for the $(2,1)$ and $(2,2)$
systems are performed using $r_0=0.01a_{\mathrm{ho}}$
while those for the $(3,2)$ and $(3,3)$
systems are performed using $r_0=0.05a_{\mathrm{ho}}$.
The radial density for the $(2,2)$ system agrees with that 
presented in 
Ref.~\protect\cite{stec08} 
after a proper rescaling
(see Ref.~\protect\cite{footnote1}).
}\label{fig_radial}
\end{figure}

\begin{figure}
\vspace*{+.9cm}
\includegraphics[angle=0,width=65mm]{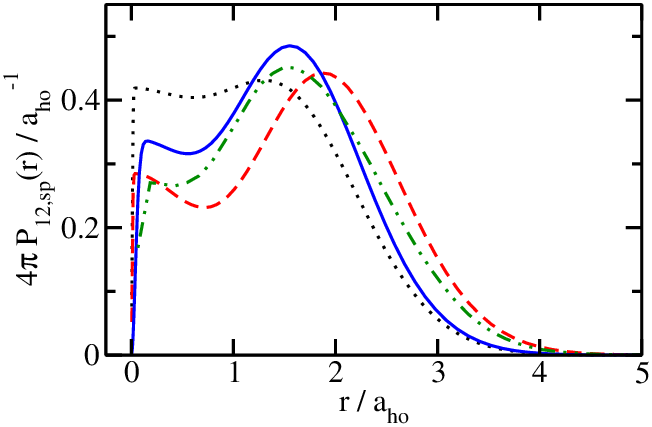}
\vspace*{0.2cm}
\caption{(Color online)
Scaled pair distribution function $4 \pi \, P_{12,\mathrm{sp}}(r)r^2$
for the lowest $L^{\Pi}=1^-$ state of
the 
$(2,1)$ system (dotted line),
the $(2,2)$ system (dashed line),
the $(3,2)$ system (solid line) and
the $(3,3)$ system (dash-dot-dotted line)
for $a_{\mathrm{ho}}/a_s=0$.
The calculations for the $(2,1)$ and $(2,2)$
systems are performed using $r_0=0.01a_{\mathrm{ho}}$
while those for the $(3,2)$ and $(3,3)$
systems are performed using $r_0=0.05a_{\mathrm{ho}}$.
The histogram bins of the $(3,3)$
system are wider than those of the
other systems, 
giving rise to the slightly different
slope of $P_{12}(r)r^2$ at small $r$.
}\label{fig_pairdist_l1}
\end{figure}

Figure~\ref{fig_radial} shows the radial densities 
$P_{1,\mathrm{sp}}(r)$ and $P_{2,\mathrm{sp}}(r)$ for 
the state with $0^+$ symmetry at unitarity
for 
the $(2,1)$ system (dotted lines),
the $(2,2)$ system (dashed lines),
the $(3,2)$ system (solid lines),
and
the $(3,3)$ system (dash-dot-dotted lines).
For the spin-balanced systems, $P_{1,\mathrm{sp}}(r)$ and $P_{2,\mathrm{sp}}(r)$
agree. 
The peak densities of the $(2,1)$, $(2,2)$ and $(3,2)$ systems 
are located at $r=0$ while the peak density of the $(3,3)$ system
is located at finite $r$. 
We interpret the fact that the peak density is either
located at $r=0$ or at finite $r$ as the system size changes as a
signature of (residual) shell structure. 
Furthermore, Fig.~\ref{fig_radial}(a) shows that the peak density of the
majority components of the $(2,1)$ and $(3,2)$
systems is smaller than that of the
$(2,2)$ system.
The minority components of the spin-imbalanced
systems, in contrast, have a higher peak density
than the $(2,2)$ system [see Fig.~\ref{fig_radial}(b)].
In interpreting the densities shown in Fig.~\ref{fig_radial}
it is important to keep in mind that the spherical components
$P_{1,\mathrm{sp}}(r)$ and $P_{2,\mathrm{sp}}(r)$ are normalized to 1. To ``account''
for the density of the entire cloud, the densities need to be 
multiplied by $N_1$ and $N_2$, respectively.

Figure~\ref{fig_pairdist_l1}
shows the scaled pair distribution function $P_{12,\mathrm{sp}}(r) r^2$
at unitarity for the lowest state with $L^{\Pi}=1^-$ symmetry.
Qualitatively, the behavior
of $P_{12,\mathrm{sp}}(r) r^2$ for the lowest states
with $L^{\Pi}=1^-$ (Fig.~\ref{fig_pairdist_l1}) and $0^+$
[Fig.~\ref{fig_pairdist_l0}(b)] at unitarity is similar,
i.e., $P_{12,\mathrm{sp}}(r)r^2$ shows a double-peak structure.
However, as already eluded to,
the scaled pair distribution functions
for the $L^{\Pi}=1^-$ state
of the spin-imbalanced systems take on vanishingly small
values at smaller $r$ values than 
those of the
spin-balanced systems.
For the $(2,1)$ 
and $(3,2)$ systems, the 
lowest $L^{\Pi}=1^-$ state has a lower energy than 
the 
lowest $0^+$ state.
Thus, a less extended and more compact pair distribution function 
for the spin-up---spin-down distance is, at
least for the systems discussed in Figs.~\ref{fig_pairdist_l0}
and \ref{fig_pairdist_l1},
associated with a lower energy.

\subsection{Non-local properties}
\label{sec_structure2}
The pair distribution functions and radial
densities discussed in the previous section
indicate that small two-component Fermi gases 
undergo significant changes as the $s$-wave scattering
length $a_s$ changes from $0^-$ over $\infty$ to $0^+$.
In the $a_s \rightarrow 0^+$ limit, 
the basic constituents of the molecular gas are pairs.
While the local structural properties provide
a great deal of insight into the formation of pairs, they provide no
information as to whether or not the pairs are
condensed. The determination of the molecular
condensate fraction is based,
as discussed in Sec.~\ref{sec_densitymatrix}, on 
the two-body density matrix that
measures the ``response'' of the system to moving a 
pair
from
one position in the trap to another position in the trap.
The one-body density matrix, in contrast, does not provide
a means to quantify the condensate fraction as it 
measures the response of the system to moving a fermionic atom
from one position in the trap to another position in the trap.
In the following, we analyze both the one-body and the two-body
density matrices.

\begin{figure}
\vspace*{+.9cm}
\includegraphics[angle=0,width=65mm]{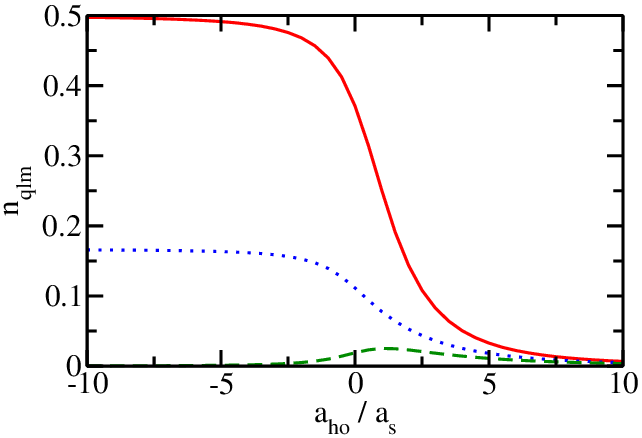}
\vspace*{0.2cm}
\caption{(Color online)
Occupation numbers $n_{qlm}$, obtained by
analyzing the one-body density matrix $\rho_1(\vec{r} \, ',\vec{r})$,
for the lowest state with $0^+$ symmetry,
i.e., the ground state, of the $(2,2)$ system
as a function of the inverse scattering length $a_s^{-1}$.
Solid, dotted and dashed lines show the
occupation numbers $n_{000}$, $n_{010}$ and $n_{100}$, respectively.
The occupation numbers $n_{011}$ and $n_{01-1}$ (not shown)
are equal to $n_{010}$.
The calculations are performed for $r_0=0.005a_{\mathrm{ho}}$.
}\label{fig_onebody_occupation}
\end{figure}

We first consider non-local
properties derived
from the one-body density matrix.
Figure~\ref{fig_onebody_occupation} shows the occupation numbers 
$n_{qlm}$ [$(qlm)=(000)$, $(100)$ and $(010)$]
for the ground state with $L^{\Pi}=0^+$ symmetry
of the $(2,2)$ system.
The behavior is similar for the $(2,1)$, $(3,2)$ and $(3,3)$ systems 
(not shown).
As $a_s$ approaches $0^-$, the numerically obtained occupation
numbers agree with the analytical results
presented in Appendix~\ref{sec_appendixc}, i.e.,
$n_{000}=1/2$, $n_{01m}=1/6$ ($m=0,\pm 1$), and $n_{qlm}=0$ 
for all other $qlm$.
These occupation numbers directly reflect the anti-symmetric
character of the 
non-interacting fermionic system: The two spin-up atoms of the 
$(2,2)$ system have to occupy different 
single-particle orbitals. One spin-up atom 
occupies the lowest harmonic oscillator orbital 
while the other spin-up atom is equally distributed among the
three degenerate first excited state harmonic oscillator 
orbitals. 
Figure~\ref{fig_onebody_occupation} shows that the occupation numbers
$n_{000}$ (solid line)
and $n_{010}$ (dotted line)
of the $(2,2)$ system change
only weakly for $a_{\mathrm{ho}}/a_{s} \lesssim -2.5$, i.e., 
the one-body density matrix $\rho_1(\vec{r} \, ',\vec{r})$
can be decomposed with fairly good accuracy
by including just four natural orbitals.
In the strongly-interacting regime, 
$n_{000}$ and $n_{010}$ decrease notably
while other occupation numbers such as $n_{100}$
(dashed line in Fig.~\ref{fig_onebody_occupation}) increase.
In this regime,
the system can no longer be thought of as a
weakly-perturbed atomic 
Fermi gas.
For $a_{\mathrm{ho}}/a_s \gtrsim 5$, 
we find that a
relatively large number of $n_{qlm}$ take on non-vanishing but small
values.
Intuitively, this can be understood as follows: An expansion
of a tight composite boson wave function in terms of effective single particle 
orbitals (the natural orbitals) requires many terms.

\begin{figure}
\vspace*{+.9cm}
\includegraphics[angle=0,width=65mm]{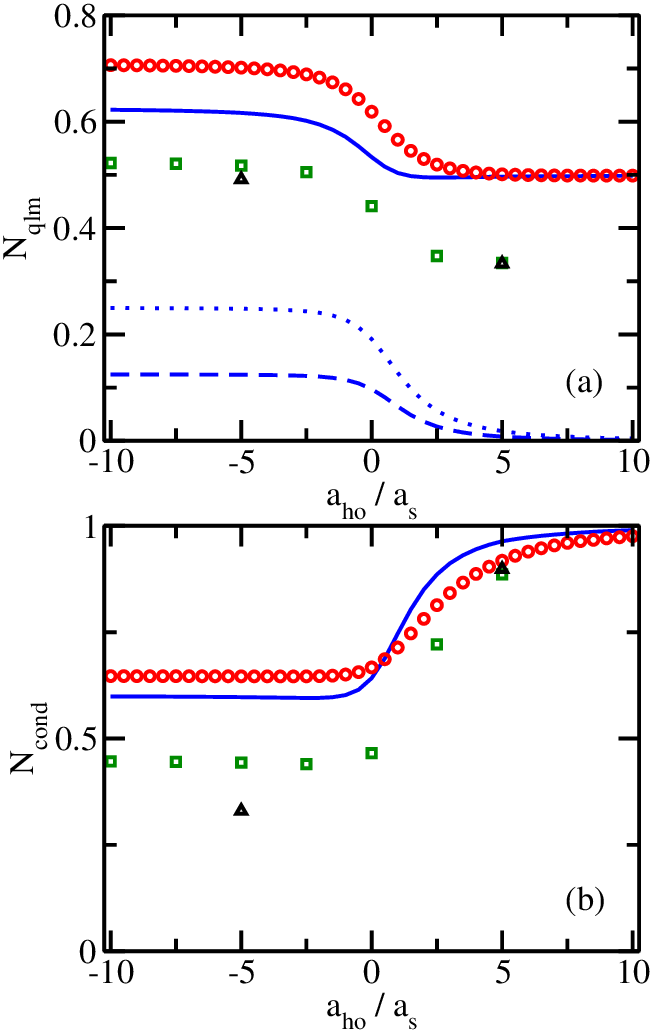}
\vspace*{0.2cm}
\caption{(Color online)
Occupation numbers $N_{qlm}$ and
condensate fraction $N_{\mathrm{cond}}$, obtained by  
analyzing the reduced two-body density matrix
$\rho_{\mathrm{red}}(\vec{R} \, ',\vec{R})$,
for the energetically lowest-lying state with $0^+$ symmetry
as a function of the inverse scattering length $a_s^{-1}$.
(a) Solid, dotted and dashed lines show the
occupation numbers $N_{000}$, $N_{100}$, and $N_{010}$, respectively,
for the $(2,2)$ system.
The occupation numbers $N_{011}$ and $N_{01-1}$ (not shown)
are equal to $N_{010}$.
For comparison, circles, squares and triangles show the
occupation number $N_{000}$ for the $(2,1)$, $(3,2)$ and $(3,3)$
systems, respectively.
(b) Circles, the solid line, squares and triangles show the 
condensate fraction $N_{\mathrm{cond}}$,
Eq.~(\ref{eq_condfracnumerical}), for the $(2,1)$, 
$(2,2)$, $(3,2)$ and $(3,3)$
systems.
The calculations are performed for 
$r_0=0.01a_{\mathrm{ho}}$ for the $(2,1)$ system,
$r_0=0.005a_{\mathrm{ho}}$ for the $(2,2)$ system, and
$r_0=0.05a_{\mathrm{ho}}$ for the $(3,2)$ and $(3,3)$ systems.
}\label{fig_twobody_occupation}
\end{figure}

\begin{figure}
\vspace*{+.9cm}
\includegraphics[angle=0,width=65mm]{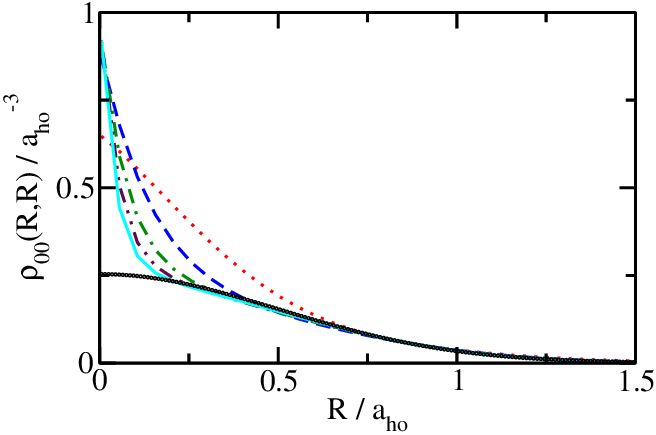}
\vspace*{0.2cm}
\caption{(Color online)
Diagonal elements $\rho_{00}(R,R)$,
obtained from the reduced
two-body density matrix $\rho_{\mathrm{red}}(\vec{R} \, ',\vec{R})$,
for
the lowest state with $L^{\Pi}=0^+$ symmetry of 
the $(2,2)$ system
for
$a_{\mathrm{ho}}/a_s=0$ (dotted line),
$a_{\mathrm{ho}}/a_s=2.5$ (dashed line),
$a_{\mathrm{ho}}/a_s=5$ (dash-dotted line),
$a_{\mathrm{ho}}/a_s=7.5$ (dash-dot-dotted line), and
$a_{\mathrm{ho}}/a_s=10$ [grey (cyan) solid line].
The calculations 
are performed for 
$r_0=0.005a_{\mathrm{ho}}$.
For comparison, the black solid line shows the quantity 
$\rho_{\mathrm{boson}}(R,R)/2$
[see discussion in the main text
and after Eq.~(\ref{eq_bosonorbital})].
}\label{fig_twobody_diag}
\end{figure}

\begin{figure}
\vspace*{+.9cm}
\includegraphics[angle=0,width=65mm]{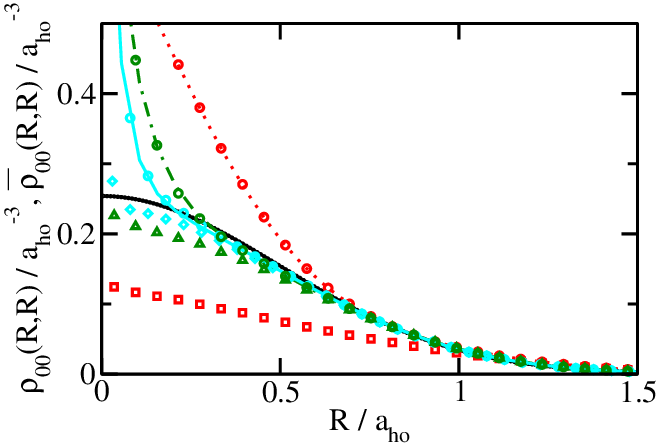}
\vspace*{0.2cm}
\caption{(Color online)
Analysis of the reduced two-body density matrix
for the lowest state with $0^+$ symmetry,
i.e., the ground state, of the $(2,2)$ system.
Dotted, dash-dotted and grey (cyan) solid lines show 
the diagonal element 
$\rho_{00}(R,R)$ obtained by a direct evaluation of
the matrix elements for $a_{\mathrm{ho}}/a_s=0$, $5$ and $10$, respectively
(these data are also shown in Fig.~\ref{fig_twobody_diag}).
For comparison, circles show $\rho_{00}(R,R)$ for the same scattering lengths
but calculated by Monte
Carlo sampling; the agreement is excellent.
Squares, triangles and diamonds 
show the diagonal element $\bar{\rho}_{00}(R,R)$ 
for $a_{\mathrm{ho}}/a_s=0$, $5$ and $10$, respectively.
The noise visible at small $R$ is a direct consequence of
the Monte
Carlo sampling aproach.
For comparison, the black
solid line shows the quantity $\rho_{\mathrm{boson}}(R,R)/2$
[see discussion in the main text and
after Eq.~(\ref{eq_bosonorbital})].
}\label{fig_montecarlo}
\end{figure}

Figures~\ref{fig_twobody_occupation} and \ref{fig_twobody_diag}
show results obtained by analyzing 
the reduced two-body density matrix $\rho_{\mathrm{red}}(\vec{R} \, ',\vec{R})$.
To aid with the interpretation of these results,
Fig.~\ref{fig_montecarlo} compares results obtained by analyzing
$\rho_{\mathrm{red}}(\vec{R} \, ',\vec{R})$
and $\bar{\rho}_{\mathrm{red}}(\vec{R} \, ',\vec{R})$, respectively;
these quantities have been introduced in the last two paragraphs of 
Sec.~\ref{sec_densitymatrix} 
to help quantify the molecular condensate fraction.
Figure~\ref{fig_twobody_occupation}(a) shows the occupation numbers 
$N_{qlm}$ for the lowest state with $0^+$ symmetry
throughout the crossover for the $(2,1)$, $(2,2)$, $(3,2)$ and $(3,3)$ systems.
For the $(2,2)$ system, e.g.,
$N_{000}$ (solid line) decreases nearly monotonically 
from $5/8$ in the $a_s \rightarrow 0^-$
limit to $1/2$ in the $a_s \rightarrow 0^+$ limit
(see Appendix~\ref{sec_appendixc});
in fact, $N_{000}$ reaches a minimum of
about $0.495$ at $a_{\mathrm{ho}}/a_s \approx 2.5$
and then 
increases again.
While it might be surprising at first sight that the occupation 
number $N_{000}$ of the lowest natural orbital is larger in the absence of pairs 
($a_s \rightarrow 0^-$ limit)
than in the 
presence of pairs ($a_s \rightarrow 0^+$ limit), 
this
is a direct consequence of the definition 
of $\rho_{\mathrm{red}}(\vec{R} \, ',\vec{R})$:
$N_{000}$ is of the order of $1/N_1$ in both limits
(see Appendix~\ref{sec_appendixc}).

The above discussion indicates that $N_{000}$ 
does not directly measure the condensate fraction of pairs.
Instead,
we call the system condensed
when the lowest natural orbital is macroscopically
occupied, i.e., when $N_{000}$ is much larger than all other
$N_{qlm}$, $(qlm) \ne (000)$.
Correspondingly, we introduce the
quantity $N_{\mathrm{cond}}$,
\begin{eqnarray}
\label{eq_condensatefraction}
N_{\mathrm{cond}}=1 - \frac{\max(\sum_{m=-l}^lN_{qlm})}{N_{000}}, \;\; (ql)\ne (00).
\end{eqnarray}
The summation over $m$ in the second term on the right hand side
of Eq.~(\ref{eq_condensatefraction}) is included since 
we could have defined the projections [see
Eq.~(\ref{eq_proj}) 
for the one-body density matrix; 
the same argument applies to the two-body density matrix]
in terms of Legendre polynomials that
depend on $l$ only instead of in terms of spherical harmonics that depend 
on $l$ and $m$.
In the $a_s \rightarrow 0^+$ limit, the second term on the right hand side
of Eq.~(\ref{eq_condensatefraction}) 
is small and $N_{\mathrm{cond}}$ approaches 1.
In the $a_s \rightarrow 0^-$ limit, 
the 
second term on the right hand side 
of Eq.~(\ref{eq_condensatefraction}) is of the order of 1
for large numbers of particles
and $N_{\mathrm{cond}}$ approaches 0.
For small systems, however,
$N_{\mathrm{cond}}$  becomes a fraction smaller than 1,
i.e., $N_{\mathrm{cond}} = 11/17, 3/5, 0.448, 1/3$ for the
non-interacting $(2,1)$,
$(2,2)$, $(3,2)$ and $(3,3)$ systems, respectively.

In practice, our analysis is limited to a finite number of
$(lm)$ projections of the reduced density matrix and 
Eq.~(\ref{eq_condensatefraction})
cannot be evaluated as is. Instead, we 
employ a slightly modified working definition of
the condensate fraction $N_{\mathrm{cond}}$,
\begin{eqnarray}
\label{eq_condfracnumerical}
N_{\mathrm{cond}}=1 - \frac{\max_q(N_{q>0,00},\sum_{m=-1}^1N_{q1m})}{N_{000}}.
\end{eqnarray}
For the systems studied in this
paper, Eqs.~(\ref{eq_condensatefraction})
and (\ref{eq_condfracnumerical}) give identical or very 
similar results.
Figure~\ref{fig_twobody_occupation}(b)
illustrates the behavior of $N_{\mathrm{cond}}$, Eq.~(\ref{eq_condfracnumerical}), 
for the 
lowest $L^{\Pi}=0^+$ state of the $(2,1)$,
$(2,2)$, $(3,2)$ and $(3,3)$ systems.
Figure~\ref{fig_twobody_occupation}(b) shows that $N_{\mathrm{cond}}$ 
increases monotonically from a finite value for
$a_{\mathrm{ho}}/a_s=-10$ 
to nearly 1 for $a_{\mathrm{ho}}/a_s=10$.
Although the quantitative behavior of $N_{\mathrm{cond}}$ 
depends on the system size, the qualitative behavior
is similar for the systems investigated.
The condensate fraction $N_{\mathrm{cond}}$
is fairly close to one for $a_{\mathrm{ho}}/a_s \gtrsim 5$.
The condensate fraction of small few-fermion
systems [Fig.~\ref{fig_twobody_occupation}(b)]
exhibits a qualitatively similar behavior to that 
of the homogeneous system~\cite{astr05}. The main
difference is that $N_{\mathrm{cond}}$ for the trapped system 
approaches, for the reasons 
discussed above, a finite value and not a vanishingly small value
as $a_s \rightarrow 0^-$.

To gain further insight into the correlations associated with the pair
formation, Fig.~\ref{fig_twobody_diag}
exemplarily shows the diagonal element $\rho_{00}(R,R)$,
obtained by analyzing the two-body density matrix,
for the ground state of the $(2,2)$ system for various scattering lengths.
For small scattering lengths ($a_s>0$), 
i.e., $a_{\mathrm{ho}}/a_s \gtrsim 2.5$,
the diagonal element $\rho_{00}(R,R)$ contains a broad 
Gaussian-like background and a sharp shorter-ranged peak.
The latter feature becomes narrower with decreasing scattering 
length. The peak falls off exponentially
and is roughly given by the square of the 
$s$-wave pair function $\Phi_{\mathrm{int}}(r)$,
Eq.~(\ref{eq_bosonpair}).
The sharp peak arises from contributions 
associated with ``large pairs'' (see also
discussion in the context of Fig.~\ref{fig_montecarlo}).
Interestingly, the sharp peak of $\rho_{00}(R',R)$
contributes negligibly to the value of $N_{000}$. This can be 
readily rationalized by realizing that the small $R'$ and $R$ 
parts of $\rho_{00}(R',R)$
are highly suppressed due to the radial volume element. 
The broad Gaussian-like peak is to a fairly good approximation
described by $\rho_{\mathrm{boson}}(R,R)/2$ 
(solid line in Fig.~\ref{fig_twobody_diag}).
The quantity $\rho_{\mathrm{boson}}(R',R)$ 
is defined in Appendix~\ref{sec_appendixc}  
after Eq.~(\ref{eq_bosonorbital}) and denotes the density matrix
for a sample of non-interacting molecules of mass $2m_a$.
In the $a_s \rightarrow 0^+$ limit,
$\rho_{\mathrm{boson}}(R,R)/2$ is expected to provide a good description.
The non-diagonal elements [i.e., $\rho_{00}(R',R)$
for $R' \ne R$, not shown] show qualitatively similar features as
the diagonal elements.
We find that
the broad background of
$\rho_{00}(R',R)$ approaches 
$\rho_{\mathrm{boson}}(R',R)/2$ 
as $a_s$ approaches the $0^+$ limit.

Figure~\ref{fig_montecarlo} compares the 
diagonal elements $\rho_{00}(R,R)$  and
$\bar{\rho}_{00}(R,R)$ of the $(2,2)$ system
for $a_{\mathrm{ho}}/a_s=0$, $5$ and $10$, respectively.
The quantity $\bar{\rho}_{00}(R,R)$, determined through Metropolis
sampling, accounts only for ``large'' distances between pairs, thereby
reflecting correlations between tightly-bound composite
molecules.
While the broad peak of $\rho_{00}(R,R)$ 
nearly coincides with $\bar{\rho}_{00}(R,R)$
for $a_{\mathrm{ho}}/a_s=10$, 
the broad peak of 
$\rho_{00}(R,R)$ 
has roughly twice as large of
an amplitude as $\bar{\rho}_{00}(R,R)$
for $a_{\mathrm{ho}}/a_s=0$.
The behavior for the non-diagonal elements, not shown, is similar
to that of the diagonal elements. 
This confirms our
interpretation above: The pairs that make up the condensate
are those with the smallest interparticle distances. 
For $a_{\mathrm{ho}}/a_s=0$, 
the
$(qlm)=(000)$ orbital is not yet exclusively occupied by the
$N_2$ smallest pairs but is occupied nearly equally
by ``small'' and ``large''
pairs. 
For $a_{\mathrm{ho}}/a_s=10$, the $(000)$ orbital is nearly exclusively occupied 
by large pairs and $\bar{\rho}_{00}(R',R) \approx \rho_{\mathrm{boson}}(R',R)/2$.
This is consistent with our finding above that the 
condensate fraction is notably smaller than 1 at unitarity.
In particular, 
a value of $N_{000} \approx 1/N_1$ at unitary does not signal 
the condensation of pairs
while a value of $N_{000} \approx 1/N_1$ in the $a_s \rightarrow 0^+$ 
limit, provided all other $N_{qlm}$ are small, does signal the
condensation of pairs.

As an alternative to Eq.~(\ref{eq_condensatefraction}), one could quantify
the condensate fraction 
in terms of the occupation number 
$\bar{N}_{000}$ associated with $\bar{\rho}_{00}(R',R)$,
i.e., $\bar{N}_{\mathrm{cond}}=N_1 \bar{N}_{000}$.
While this might be, in certain respects, a more intuitive measure
than Eq.~(\ref{eq_condensatefraction}),
the determination
of $\bar{\rho}_{00}(R',R)$ and thus $\bar{N}_{000}$ is, within our framework,
computationally significantly more involved than that
of ${\rho}_{00}(R',R)$. Thus, we did not apply this alternative measure.

\begin{figure}
\vspace*{+.9cm}
\includegraphics[angle=0,width=70mm]{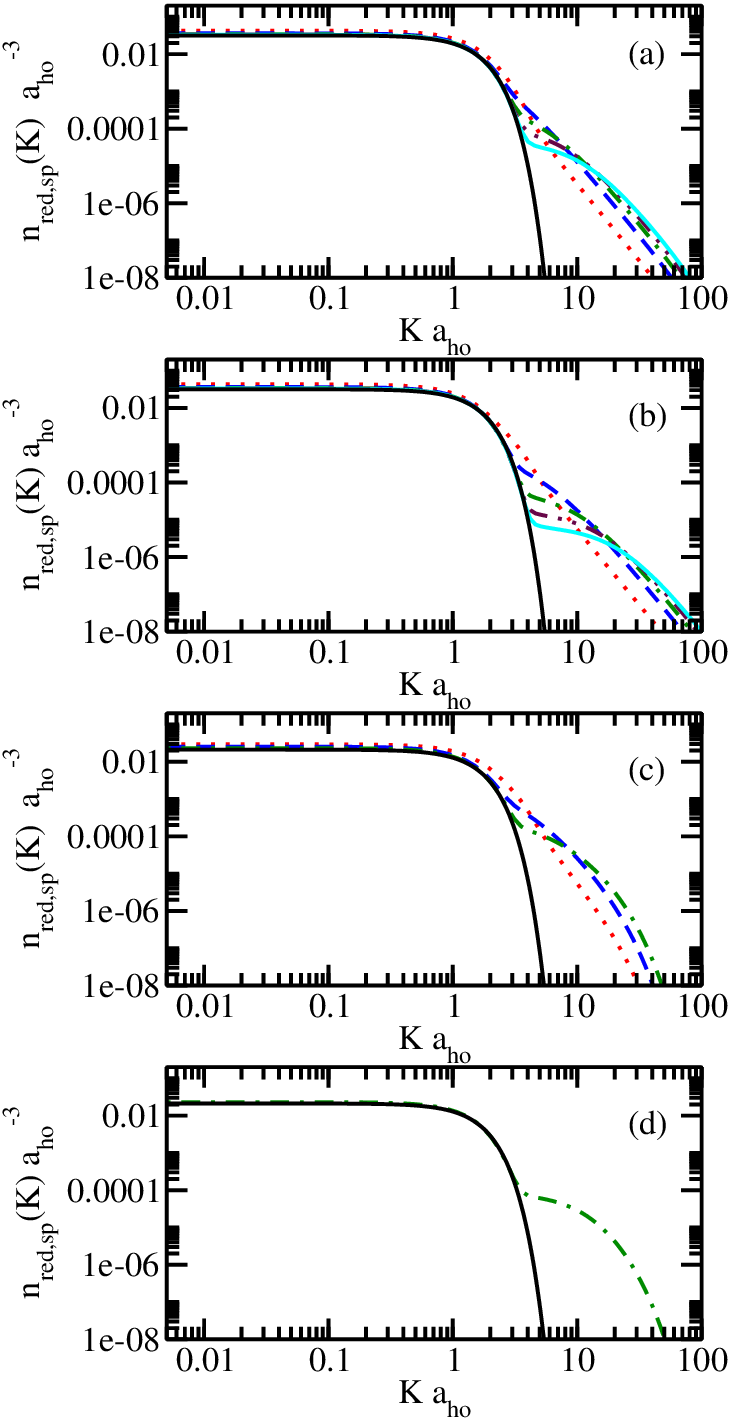}
\vspace*{0.2cm}
\caption{(Color online)
Momentum distribution $n_{\mathrm{red},\mathrm{sp}}(K)$ for 
the lowest state with $L^{\Pi}=0^+$ symmetry for
(a) the $(2,1)$ system,
(b) the $(2,2)$ system,
(c) the $(3,2)$ system, and
(d) the $(3,3)$ system.
Dotted, dashed, dash-dotted, dash-dot-dotted and
grey (cyan) solid lines
are for $a_{\mathrm{ho}}/a_s=0$, $2.5$, $5$, $7.5$ and $10$, respectively
(for $N=5$, the largest $a_{\mathrm{ho}}/a_s$ considered is $5$;
for $N=6$, results are shown for $a_{\mathrm{ho}}/a_s=5$ only).
For comparison, the dark solid lines
show the quantity $n_{\mathrm{boson},\mathrm{sp}}(K)/N_1$,
Eq.~(\protect\ref{eq_momentumboson}) [or first term on the
right hand side of Eq.~(\ref{eq_momentumanalytical})].
The calculations for the $(2,1)$, $(2,2)$,
$(3,2)$ and $(3,3)$
systems are performed using $r_0=0.01a_{\mathrm{ho}}$, $r_0=0.005a_{\mathrm{ho}}$,
$r_0=0.05a_{\mathrm{ho}}$ and 
$r_0=0.05a_{\mathrm{ho}}$, respectively.
Note the log-log scale.
}\label{fig_momentumpair}
\end{figure}

Lastly, we consider the 
momentum distribution $n_{\mathrm{red},\mathrm{sp}}(K)$
associated with the center-of-mass vector of spin-up---spin-down
pairs.
Figure~\ref{fig_momentumpair}
shows that $n_{\mathrm{red},\mathrm{sp}}(K)$
consists of two parts, a feature at smaller $K$ 
($K \lesssim 5a_{\mathrm{ho}}^{-1}$)
and a feature that extends
to much larger $K$ values.
The emergence of these two features with decreasing $a_s$
is another indication of the
condensation of pairs.
The small and large $K$ features become more distinctly separated 
as $a_s$ decreases. This is in agreement with the 
increase of $N_{\mathrm{cond}}$ with decreasing $a_s$.
In fact, Fig.~\ref{fig_momentumpair} suggests that 
the few-fermion system can be called condensed when 
the momentum distribution $n_{\mathrm{red},\mathrm{sp}}(K)$ shows two clearly
distinguishable features, i.e., when the derivative of $n_{\mathrm{red},\mathrm{sp}}(K)$
exhibits a significant change for a small change in $K$.

\begin{figure}
\vspace*{+.9cm}
\includegraphics[angle=0,width=70mm]{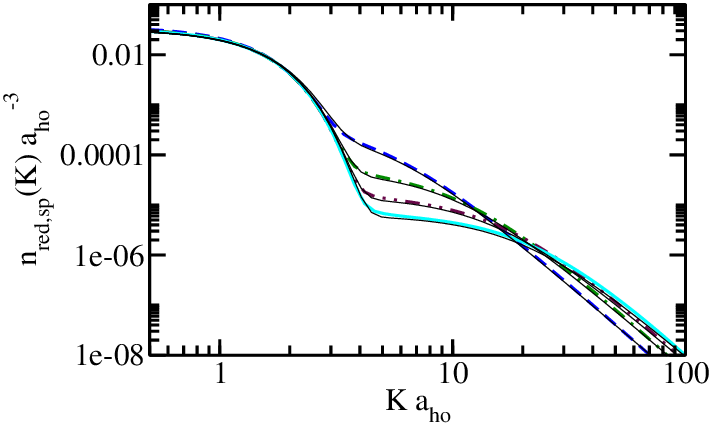}
\vspace*{0.2cm}
\caption{(Color online)
Momentum distribution $n_{\mathrm{red},\mathrm{sp}}(K)$ for 
the lowest state with $L^{\Pi}=0^+$ symmetry of
the $(2,2)$ system.
Dashed, dash-dotted, dash-dot-dotted and
grey (cyan) solid lines
are for $a_{\mathrm{ho}}/a_s=2.5$, $5$, $7.5$ and $10$, respectively
[these data are also shown in Fig.~\ref{fig_momentumpair}(b)].
For comparison, the thin solid lines
show the analytically predicted momentum distribution
$n_{\mathrm{red,sp}}(K)$, Eq.~(\ref{eq_momentumanalytical}).
Note the log-log scale.
}\label{fig_momentumpairreferee}
\end{figure}

In the $a_s \rightarrow 0^+$ limit,
the momentum distribution $n_{\mathrm{red,sp}}(K)$
for systems with $N_1=N_2$ 
is well described by the analytical expression
(see Appendix~\ref{sec_appendixc})
\begin{eqnarray}
\label{eq_momentumanalytical}
n_{\mathrm{red,sp}}(K) \approx
\frac{1}{N_1}
\frac{a_{\mathrm{ho}}^3}
{{(2\pi)}^{3/2}}
\exp \left( - (a_{\mathrm{ho}} {K})^2 /2\right)
+ \nonumber \\
\frac{N_1-1}{N_1} 
\frac{a_{\mathrm{ho}}^3}{\pi^{3/2} K a_{\mathrm{ho}}} 
\Re \left[ -i A \exp(A^2) \mathrm{Erfc}(A) \right]
 ,
\end{eqnarray}
where $A=2 a_{\mathrm{ho}}/a_s + i K a_{\mathrm{ho}}$ 
and $\Re$ and $\mathrm{Erfc}$ denote
the real part and the complementary error function, respectively.
The first term on the 
right hand side of Eq.~(\ref{eq_momentumanalytical})
accounts for 
the small $K$ feature of $n_{\mathrm{red,sp}}(K)$ and represents
the momentum distribution 
$n_{\mathrm{boson},\mathrm{sp}}(K)/N_1$, 
Eq.~(\ref{eq_momentumboson}), derived
for non-interacting composite bosons of mass $2m_a$
(dark solid lines in Fig.~\ref{fig_momentumpair}).
The second term on the 
right hand side of Eq.~(\ref{eq_momentumanalytical})
accounts for 
the large $K$ feature of $n_\mathrm{red,sp}(K)$ and is associated
with the internal structure
of the composite bosons. In the large $K$ limit, the second term behaves,
as expected, 
as $1/K^4$~\cite{tan08a,tan08b,tan08c}.
The dependence of the large $K$ part of the momentum
distribution on the $s$-wave scattering length $a_s$
for systems with $N_1=N_2$
is reproduced quite
accurately
by Eq.~(\ref{eq_momentumanalytical}). 
This is illustrated 
exemplarily for the ground state
of the $(2,2)$ system in Fig.~\ref{fig_momentumpairreferee},
which compares the momentum distribution
given by Eq.~(\ref{eq_momentumanalytical}) (thin solid lines)
with
the numerically determined $n_{\mathrm{red,sp}}(K)$ for $a_{ho}/a_s=2.5-10$
[same data as shown in Fig.~\ref{fig_momentumpair}(b)].
For $K \gtrsim r_0^{-1}$ (not shown in Fig.~\ref{fig_momentumpairreferee}),
the momentum distribution given in Eq.~(\ref{eq_momentumanalytical})
deviates from that obtained numerically for finite-range interactions. This
is expected, since this is the regime where the details of the
two-body interaction potential become relevant.
The analytical expression
for $n_{\mathrm{red,sp}}(K)$ for systems with
$N_1-N_2=1$ and 
$a_s \rightarrow 0^+$ differs from Eq.~(\ref{eq_momentumanalytical})
and is given in Appendix~\ref{sec_appendixc}, Eq.~(\ref{eq_momentumspare}).

\section{Conclusions}
\label{sec_conclusions}

This paper considers small two-component Fermi gases under external spherically
symmetric confinement. 
We have treated systems with up to $N=6$ atoms, where $N_1-N_2=0$ or 1,
within a microscopic, non-perturbative zero-temperature
framework. Using the stochastic variational approach, we 
have investigated
the energetics and structural properties
as functions of the $s$-wave scattering length $a_s$
and the symmetry of the system. In certain cases,
we have also examined the dependence of the results on the range
$r_0$
of the underlying two-body
model potential.

Our analysis of the energetics and the structural properties 
extends previous studies and adds to the rapidly
growing body of results for small trapped 
three-dimensional few-fermion systems.
In particular, we have presented extrapolated zero-range energies for 
the natural parity states of
the five- and six-particle systems at unitarity
for various angular momenta. These energies are
expected to serve as benchmarks for other numerical approaches.

We have also presented a detailed study of the non-local properties
of few-fermion systems. One of our goals has been to
quantify the molecular condensate fraction
of trapped two-component Fermi systems on the positive scattering
length side. To this end, we have analyzed the one-body and the 
two-body density matrices and proposed to use
the quantity $N_{\mathrm{cond}}$ as a measure of the
molecular condensate fraction.
We showed that 
the momentum distribution $n_{\mathrm{red,sp}}(K)$,
an experimentally accessible observable, develops
two clearly
distinguishable features
at 
$s$-wave scattering lengths $a_s$ 
for which the molecular condensate 
fraction $N_{\mathrm{cond}}$ takes on values close to 
$1$. 

The determination of the molecular condensate fraction of the
trapped system is more complicated than that of the
homogeneous Fermi system since the trap ``cuts off'' the asymptotic behavior
that is typically analyzed to determine the molecular
condensate fraction
of the homogeneous system
(see, e.g., Ref.~\cite{astr05} for a cold-atom study). 
Instead, the analysis 
of finite-sized systems proceeds through the diagonalization
of the two-body
density matrix. The diagonalization results in a set of
natural orbitals and occupation numbers that can then be
used to quantify the molecular condensate fraction. 
In our approach, we measured the position vectors of the
composite pairs with respect to the trap center. Alternatively, one might
imagine measuring the position vectors with respect to the 
center of mass of the trapped system.
In the context of bosonic systems, implications of 
defining the one-body density matrix in terms
of different ``reference coordinates''
have been discussed in the literature~\cite{peth00,thog07,yama08,yama09}.
Future work needs to address how 
the results obtained by analyzing the two-body density matrix
of fermionic few-body systems
depend on the use of
different reference coordinates.

\appendix

\section{Matrix elements employed in stochastic variational
approach}
\label{sec_appendixa}
While explicit expressions for the Hamiltonian and overlap matrix
elements are available in the literature~\cite{cgbook},
explicit expressions 
for the non-local
observables that we are interested in are not.
Thus,
this appendix outlines the derivation of selected 
matrix elements used in our SV calculations; our derivations 
follow the general approach outlined in Ref.~\cite{cgbook}. 

In
our implementation, we construct the basis set by treating the
relative Jacobi vectors $\vec{\rho}_1,\cdots,\vec{\rho}_{N-1}$
only. The
structural properties,
however, are determined by multiplying the optimized basis set by the
unnormalized ground state center-of-mass wave function 
$\psi_{\mathrm{cm}}(\vec{R}_{\mathrm{cm}})$ [Eq.~(\ref{eq_cm}) with $N_{\mathrm{cm}}=1$].
The unsymmetrized (and unnormalized) basis functions 
$\phi_A$ that include the center-of-mass
degrees of freedom and describe states with $L^{\Pi}=0^+$
symmetry read
\begin{eqnarray}
\label{eq_cgbasis}
\phi_A(\vec{x}) = \exp \left( -\frac{1}{2} \vec{x}^T \, \underline{A} \, \vec{x} \right),
\end{eqnarray}
where $\vec{x}$ collectively denotes the $N$ Jacobi vectors,
$\vec{x}=(\vec{\rho}_1,\cdots,\vec{\rho}_{N-1},\vec{R}_{\mathrm{cm}})$.
Here, $\underline{A}$ is a symmetric and positive definite
matrix that is written in terms of
$(N-1)(N-2)/2$ variational parameters [the $(\underline{A})_{jk}$ with
$j=1,\cdots,N-1$ and $k \ge j$ are optimized semi-stochastically].
To ensure that the center-of-mass degrees of freedom 
are in the ground state, the matrix elements
$(\underline{A})_{jN}$ and $(\underline{A})_{Nj}$, where $j=1,\cdots,N-1$
are set to zero and the matrix element $(\underline{A})_{NN}$
is set to $N/a_{\mathrm{ho}}^2$.
The Jacobi vectors $\vec{x}$ and the single particle
coordinates $\vec{y}=(\vec{r}_1,\cdots,\vec{r}_N)$
are related through the $N \times N$ matrix $\underline{U}$,
\begin{eqnarray}
\vec{x}=\underline{U} \, \vec{y}.
\end{eqnarray} 

Our first goal is to determine the matrix element 
$(\rho_{1}(\vec{r} \, ',\vec{r}))_{A'A}=
\langle \phi_{A'}| \rho_{1} |\phi_A \rangle/
\langle \phi_{A'}| \phi_A \rangle$,
\begin{eqnarray}
\label{eq_a2}
(\rho_{1}(\vec{r} \, ',\vec{r}))_{A'A}= (O_{A'A})^{-1}
\nonumber \\
\times
\int 
\left[ \int \delta(\vec{r} \, '-\vec{r}_1) \phi_{A'}(\vec{x}) d^3 \vec{r}_1 \right] 
\nonumber \\
\times \left[ \int \delta(\vec{r}-\vec{r}_1) \phi_{A}(\vec{x}) d^3 \vec{r}_1 \right] 
d^{3N-3} \vec{y}_{\mathrm{red}},
\end{eqnarray}
where
$\vec{y}_{\mathrm{red}}=(\vec{r}_2,\cdots,\vec{r}_N)$
and
\begin{eqnarray}
O_{A'A}= 
\left(
\frac{(2 \pi)^{N}}{\det( \underline{A}'+\underline{A})}\right)^{3/2}.
\end{eqnarray}
It is convenient~\cite{cgbook}
to rewrite the right hand side of
Eq.~(\ref{eq_a2}) in terms of the function $g(\vec{s};\underline{A},\vec{x})$,
\begin{eqnarray}
g(\vec{s};\underline{A},\vec{x})=
\exp \left( -\frac{1}{2} \vec{x}^T \, \underline{A} \, \vec{x} +
\vec{s}^T \vec{x} \right),
\end{eqnarray}
where $\vec{s}$ denotes a vector that has the same dimensionality as
$\vec{x}$.
The unsymmetrized basis functions can then be written as
$\phi_A(\vec{x}) = g(0;\underline{A},\vec{x})$.
Using that $\vec{x}^T \, \underline{A} \, \vec{x}=
\vec{y}^T \, \underline{U}^T \, \underline{A} \, \underline{U} \,
\vec{y}$, we 
rewrite the unsymmetrized basis functions $\phi_A$ in terms of 
$\vec{y}$  and separate off the
$\vec{r}_1$ dependence,
\begin{eqnarray}
\label{eq_a2a}
\phi_A(\vec{y})=g(0;\underline{B},\vec{y}_{\mathrm{red}})
\exp \left( 
-\frac{1}{2} b_1 \vec{r}_1^2 - (\vec{b}^T \vec{y}_{\mathrm{red}})^T \vec{r}_1
\right).
\end{eqnarray}
Here, the scalar $b_1$
is given by
$(\underline{U}^T \, \underline{A} \, \underline{U})_{11}$,
the $(N-1)$-dimensional vector
 $\vec{b}$ is given
by
$((\underline{U}^T \, \underline{A} \, \underline{U})_{12},\cdots,
(\underline{U}^T \, \underline{A} \, \underline{U})_{1N})$,
 and the $(N-1)\times (N-1)$-dimensional
matrix $\underline{B}$ 
is given by 
$\underline{U}^T \, \underline{A} \, \underline{U}$ with the first row and
column removed.
In Eq.~(\ref{eq_a2a}), the quantity
$(\vec{b}^T \vec{y}_{\mathrm{red}})^T \vec{r}_1$
equals $\sum_{j=2}^N (\vec{b})_{j-1} \vec{y}_j^T \, \vec{r}_1$,
where $(\vec{b})_j$ denotes the $j$th element of the vector $\vec{b}$.
To evaluate the right hand side of Eq.~(\ref{eq_a2}),
we 
define $b_1'$, $\vec{b} \, '$ and $\underline{B}'$ analogously to
$b_1$, $\vec{b}$ and $\underline{B}$.
This yields
\begin{eqnarray}
\label{eq_a3}
(\rho_{1}(\vec{r} \, ',\vec{r}))_{A'A}=
(O_{A'A})^{-1}
\int 
g(0;\underline{B}',\vec{y}_{\mathrm{red}})
g(0;\underline{B},\vec{y}_{\mathrm{red}}) \nonumber\\
\times
\exp\left( -\frac{1}{2} b_1' \vec{r} \, '^2 - (\vec{b} \,' \, ^T \vec{y}_{\mathrm{red}})^T \vec{r} \, '
\right) \nonumber \\
\times
\exp\left( -\frac{1}{2} b_1 \vec{r}^2 - (\vec{b}^T \vec{y}_{\mathrm{red}})^T \vec{r}
\right)
d^{3N-3} \vec{y}_{\mathrm{red}},
\end{eqnarray}
which can be rewritten as
\begin{eqnarray}
\label{eq_a4}
(\rho_{1}(\vec{r} \, ',\vec{r}))_{A'A}=
(O_{A'A})^{-1}
\int
\exp \left( -\frac{1}{2} (b_1' \vec{r} \, ' \, ^2 + b_1 \vec{r}^2) \right) \nonumber \\
g(-(\vec{b} \, ' \vec{r} \, ' +\vec{b} \vec{r});\underline{B}' + \underline{B},\vec{y}_{\mathrm{red}})
d^{3N-3} \vec{y}_{\mathrm{red}}.
\end{eqnarray}
Here, the quantity $\vec{b} \vec{r}$
is a $(N-1)$-dimensional
vector with elements $(\vec{b})_j \vec{r}$, where $j=1,\cdots,N-1$.
Using the first entry of Table 7.1 of Ref.~\cite{cgbook},
\begin{eqnarray}
\int g(\vec{s};\underline{A},\vec{x}) d^{3N} \vec{x}=
\left( \frac{(2 \pi)^N}{\det{\underline{A}}} \right)^{3/2}
\exp \left( \frac{1}{2} \vec{s}^T \, \underline{A}^{-1} \, \vec{s} \right),
\end{eqnarray}
we find a compact expression for the
matrix elements of the one-body density matrix,
\begin{eqnarray}
\label{eq_a5}
(\rho_{1}(\vec{r} \, ',\vec{r}))_{A'A}=
\nonumber \\
(O_{A'A})^{-1}
c_1
\exp \left(
-\frac{c'}{2}  \vec{r} \, ' \, ^2
-\frac{c}{2}  \vec{r}^2
+ \frac{a}{2} \vec{r} \, ' \, ^T \vec{r}
\right),
\end{eqnarray}
where 
\begin{eqnarray}
c_1=\left( \frac{( 2 \pi)^{N-1}}{\mathrm{det} ( \underline{B}'+\underline{B})}\right)^{3/2},
\end{eqnarray}
\begin{eqnarray}
c' = b_1'-\vec{b} \, ' \, ^T \, \underline{C} \, \vec{b} \, ',
\end{eqnarray}
\begin{eqnarray}
c = b_1-\vec{b}^T \, \underline{C} \, \vec{b},
\end{eqnarray}
\begin{eqnarray}
a=\vec{b} \, ' \, ^T \, \underline{C} \, \vec{b} + 
\vec{b}^T \, \underline{C} \, \vec{b} \, ',
\end{eqnarray}
and
\begin{eqnarray}
\underline{C}=(\underline{B}'+\underline{B})^{-1}.
\end{eqnarray}

We now use
Eq.~(\ref{eq_a5}) to determine an analytical expresssion
for the matrix element $(\rho_{00}(r',r))_{A'A}$.
To this end, we write $\vec{r} \, ' \, ^T \vec{r}=r' r \cos \gamma$,
where $\gamma$ denotes the angle between $\vec{r} \, '$ and $\vec{r}$.
The integration over $\theta$, $\varphi$, $\theta'$ and $\varphi'$
then reduces to a single integration over $\gamma$ (the other integrations
give a factor of $8 \pi^2$).
Performing the integration over $\gamma$ yields
\begin{eqnarray}
(\rho_{00}({r}',{r}))_{A'A}= \nonumber \\
(O_{A'A})^{-1}
\frac{2 c_1}{a r' r} \exp \left[-\frac{1}{2}(c' r'^2+c r^2) \right] 
\sinh \left( \frac{a r r'}{2} \right).
\end{eqnarray}
The matrix elements for 
higher partial wave projections can be determined in a similar manner.

Our next goal is to determine an 
analytical expression for the
matrix element $(n_{1,\mathrm{sp}}({k}))_{A'A}$.
Using Eqs.~(\ref{eq_momentumdist}) 
and (\ref{eq_a5}), we write
\begin{eqnarray}
\label{eq_a10}
(n_{1}(\vec{k}))_{A'A}=
(O_{A'A})^{-1} \nonumber \\
\times 
\frac{c_1}{(2 \pi)^3} \int  
\exp\left[-\frac{1}{2}(c' r'^2 + c r^2 - a \vec{r} \, ' \, ^T \vec{r} )\right]
\nonumber \\
\times \exp[i \vec{k}^T (\vec{r} \, '-\vec{r})]
d^3 \vec{r} \, ' d^3 \vec{r}.
\end{eqnarray}
Defining $\vec{X}=\vec{r} \, '-\vec{r}$,
Eq.~(\ref{eq_a10}) becomes
\begin{eqnarray}
(n_{1}(\vec{k}))_{A'A}=
(O_{A'A})^{-1} \nonumber \\
\times
\frac{c_1}{(2 \pi)^3} \int 
\exp \left(- \frac{f}{2} r^2  - \frac{c'}{2}X^2 + \frac{g}{2}\vec{X}^T \vec{r}
\right) \nonumber \\
\times \exp(i \vec{k}^T \vec{X})
d^3 \vec{r} d^3 \vec{X},
\end{eqnarray}
where
$f = c'+c-a$
and
$g=a - 2 c'$.
Next, we expand the quantity $\exp(i \vec{k}^T \vec{X})$,
\begin{eqnarray}
\exp(i \vec{k}^T \vec{X}) =
\sum_{l=0}^{\infty} (2l+1) i^l j_l(k X) P_l(\cos \gamma'), 
\end{eqnarray}
where $\gamma'$ denotes the angle between $\vec{k}$ and $\vec{X}$.
Considering the $l=0$ component only,
we find
\begin{eqnarray}
(n_{1,\mathrm{sp}}({k}))_{A'A}=
(O_{A'A})^{-1}
\frac{c_1}{\pi} \nonumber \\
\int_0^{\infty} \int_0^{\infty} \int_{-1}^{1} 
\exp \left(- \frac{f}{2} r^2  - \frac{c'}{2}X^2 + 
\frac{g}{2} X r \cos \gamma \right)
\nonumber \\
\frac{\sin(k X)}{k X} r^2 X^2 d \cos \gamma d r dX, 
\end{eqnarray}
where $\gamma$ denotes the angle between $\vec{r}$ and $\vec{X}$.
The integration over $\cos \gamma$ gives
\begin{eqnarray}
(n_{1,\mathrm{sp}}({k}))_{A'A}=
(O_{A'A})^{-1} \nonumber \\
\times
\frac{4c_1}{\pi g k} 
\int_0^{\infty} \int_0^{\infty} 
\exp \left(- \frac{f}{2} r^2  - \frac{c'}{2}X^2  \right)
\nonumber \\
\sinh \left(\frac{g r X}{2} \right)
\sin(k X) r d r dX.
\end{eqnarray}
If $f>0$, the integration over $r$ can also be performed 
analytically,
\begin{eqnarray}
(n_{1,\mathrm{sp}}({k}))_{A'A}=
\nonumber \\
(O_{A'A})^{-1}
\frac{\sqrt{2}c_1}{\sqrt{\pi} f^{3/2} k} 
\int_0^{\infty} 
\exp \left(- \frac{d}{2}X^2  \right)
\sin(k X) X dX,
\end{eqnarray}
where $d$ is given by
\begin{eqnarray}
d = c' - \frac{g^2}{4f}.
\end{eqnarray}
Lastly, the integration over $X$ gives for $d>0$,
\begin{eqnarray}
(n_{1,\mathrm{sp}}({k}))_{A'A}=
(O_{A'A})^{-1}
\frac{c_1}{(d f) ^{3/2} } \exp \left( - \frac{k^2}{2 d} 
\right).
\end{eqnarray}
We have checked numerically that $f$ and $d$ are, indeed,
greater than 0.

With one minor change, 
the derivation outlined above for the matrix elements of
the one-body density matrix $\rho_1(\vec{r} \, ',\vec{r})$
also applies 
to the matrix elements of the reduced two-body density matrix
$\rho_{\mathrm{red}}(\vec{R} \, ',\vec{R})$.
In particular, the single-particle coordinate vector $\vec{y}$
needs to be replaced by 
$(\vec{R},\vec{r}, \vec{r}_2,\cdots,\vec{r}_{N_1},\vec{r}_{N_1+2},\cdots,\vec{r}_N)$
and the matrix $\underline{U}$ needs to be redefined accordingly.
The derivation of the matrix elements for the
quantities $\rho_{00}(R',R)$ and 
$n_{\mathrm{red},\mathrm{sp}}(K)$ then carries over without additional changes.

\section{Monte Carlo sampling of 
density matrix and momentum distribution}
\label{sec_appendixb}
This appendix discusses the determination of
various observables through the Monte Carlo sampling of the 
wave function $\psi_{\mathrm{tot}}$. 
Although our approach follows standard 
procedures~\cite{montecarlo,lewa88,dubo01},
we find it useful
to summarize a few key 
results in this appendix for completeness.

Throughout this appendix, we assume that $\psi_{\mathrm{tot}}$
is known but not necessarily normalized.
We use a Metropolis walk to generate a set of configurations
$(\vec{r}_{1,j},\cdots,\vec{r}_{N,j})$,
where
$j=1,\cdots,N_{\mathrm{sample}}$, 
that are distributed according to the probability distribution
$P(\vec{r}_1,\cdots,\vec{r}_N)$,
\begin{eqnarray}
P(\vec{r}_1,\cdots,\vec{r}_N) =  \nonumber \\
\frac{|\psi_{\mathrm{tot}}(\vec{r}_1,\cdots,\vec{r}_N)|^2}
{\int |\psi_{\mathrm{tot}}(\vec{r}_1,\cdots,\vec{r}_N)|^2 
d^3\vec{r}_1 \cdots d^3\vec{r}_N}.
\end{eqnarray}
Quite generally, the strategy is to express the
expectation value of the observable $A$ 
in terms of $P(\vec{r}_1,\cdots,\vec{r}_N)$
and an ``auxiliary function'' $A'$,
\begin{eqnarray}
\label{eq_aprimegeneral}
\langle A \rangle =
\int P(\vec{r}_1,\cdots,\vec{r}_N) A'(\vec{r}_1,\cdots,\vec{r}_N)
d^3 \vec{r}_1 \cdots d^3 \vec{r}_N,
\end{eqnarray}
and to then average the quantity $A'$ over the 
configurations generated by the Metropolis walk,
\begin{eqnarray}
\langle A \rangle = \frac{1}{N_{\mathrm{sample}}}
\sum_{j=1}^{N_{\mathrm{sample}}} A'(\vec{r}_{1,j},\cdots,\vec{r}_{N,j}).
\end{eqnarray}
The functional form of the auxiliary function $A'$ 
depends on the observable $A$ of interest. In general,
$A'$ can depend on one or more of the coordinate vectors
$\vec{r}_i$, where $i=1,\cdots,N$.
As an example, we consider the radial density $P_{1,\mathrm{sp}}(r)$,
Eq.~(\ref{eq_radialspherical}), which can be 
rewritten as
\begin{eqnarray}
P_{1,\mathrm{sp}}(r)=
\int P(\vec{r}_1,\cdots,\vec{r}_N) \frac{\delta(r-r_1)}{4 \pi r_1^2}
d^3 \vec{r}_1 \cdots d^3 \vec{r}_N.
\end{eqnarray}
We thus have
$A'=A'(r, r_1)=\delta(r-r_1)/(4 \pi r_1^2)$.

We apply an analogous strategy to 
calculate the non-local observables
$\rho_{00}(r',r)$ and $n_{1,\mathrm{sp}}(k)$.
The projected one-body density matrix $\rho_{00}({r}',{r})$,
Eq.~(\ref{eq_proj}),
can be rewritten as 
\begin{eqnarray}
\rho_{00}({r}',{r})=
\int P(\vec{r}_1,\cdots,\vec{r}_N)
\frac{1}{4 \pi}
\nonumber \\
\times
\frac{\psi_{\mathrm{tot}}^*(\vec{r} \, ',\vec{r}_2,\cdots,\vec{r}_N)}
{\psi_{\mathrm{tot}}^*(\vec{r}_1,\vec{r}_2,\cdots,\vec{r}_N)}
\frac{\delta({r}-{r}_1)}{4 \pi r_1^2} d^2 \Omega_{r'}\,
d^3 \vec{r}_1 \cdots d^3\vec{r}_N.
\end{eqnarray}
Comparison with Eq.~(\ref{eq_aprimegeneral}) shows
that the auxiliary function $A'$ 
now contains an
integration over $\hat{r} \, '$.
This integration is performed by
generating 
a unit vector $\hat{r} \, '$
with random direction for each 
configuration $(\vec{r}_{1,j},\cdots,\vec{r}_{N,j})$.
The random unit vector is then scaled to the desired
length $r'$---in our calculations we
employ a linear grid---and the 
$(r',r)$ bin of the $\rho_{00}$
histogram is increased by 
$\psi_{\mathrm{tot}}^*(\vec{r} \, ',\vec{r}_2,\cdots,\vec{r}_N)/
[\psi_{\mathrm{tot}}^*(\vec{r}_1,\vec{r}_2,\cdots,\vec{r}_N)16 \pi^2 r_1^2]$.
At the end of the sampling, we symmetrize the
projected one-body density matrix.

The Metropolis sampling of the spherical component 
$n_{1,\mathrm{sp}}(k)$ of the momentum distribution proceeds similarly
to that of $\rho_{00}(r',r)$.
In particular, we rewrite Eq.~(\ref{eq_momentumspherical}),
\begin{eqnarray}
n_{1,\mathrm{sp}}({k})=
\frac{1}{(2 \pi)^3} \nonumber \\
\times \int P(\vec{r}_1,\cdots,\vec{r}_N) 
\frac{\psi_{\mathrm{tot}}^*(\vec{r}_1+\vec{X},\vec{r}_2,\cdots,\vec{r}_N)}
{\psi_{\mathrm{tot}}^*(\vec{r}_1,\vec{r}_2,\cdots,\vec{r}_N)} 
\nonumber \\
\times
\frac{\sin(k X)}{k X}
d^3\vec{X} \, d^3\vec{r}_1 \cdots d^3\vec{r}_N.
\end{eqnarray}
The integration over $\vec{X}$ is performed in two steps.
The angular integrations are performed, as discussed above for
$\rho_{00}(r',r)$, by generating a unit vector 
$\hat{X}$ with random direction for each configuration
$(\vec{r}_{1,j},\cdots,\vec{r}_{N,j})$.
The radial integration, in turn, is performed by defining a linear grid in
$X$ and by employing the trapezoidal rule.

The Monte Carlo sampling of the quantities $\rho_{00}(R',R)$ and 
$n_{\mathrm{red},\mathrm{sp}}(K)$ proceeds analogously:
$\rho_{00}(R',R)$ and $n_{\mathrm{red},\mathrm{sp}}(K)$ are rewritten
as
\begin{eqnarray}
\rho_{00}({R}',{R})=
\int P(\vec{r}_1,\cdots,\vec{r}_N)
\frac{1}{4 \pi}
\nonumber \\
\times
\frac{\psi_{\mathrm{tot}}^*(\vec{R} \, '+\frac{1}{2}\vec{r}_1-\frac{1}{2}\vec{r}_2,
\vec{R} \, '-\frac{1}{2}\vec{r}_1+\frac{1}{2}\vec{r}_2,\vec{r}_3,\cdots,\vec{r}_N)}
{\psi_{\mathrm{tot}}^*(\vec{r}_1,\vec{r}_2,\cdots,\vec{r}_N)} 
\nonumber \\
\times
\frac{\delta\left(R - \left|\frac{\vec{r}_1+\vec{r}_2}{2}\right|\right)}
{\pi \left|\vec{r}_1+\vec{r}_2\right|^2} 
d^2 \Omega_{R'}\,
d^3 \vec{r}_1 \cdots d^3\vec{r}_N
\end{eqnarray}
and 
\begin{eqnarray}
n_{\mathrm{red},\mathrm{sp}}({K})=
\frac{1}{(2 \pi)^3} 
\int P(\vec{r}_1,\cdots,\vec{r}_N) 
\nonumber \\
\times 
\frac{\psi_{\mathrm{tot}}^*(\vec{r}_1+\vec{X},\vec{r}_2+\vec{X},\vec{r}_3,\cdots,\vec{r}_N)}
{\psi_{\mathrm{tot}}^*(\vec{r}_1,\vec{r}_2,\cdots,\vec{r}_N)} 
\nonumber \\
\times
\frac{\sin(K X)}{K X}
d^3\vec{X} \, d^3\vec{r}_1 \cdots d^3\vec{r}_N,
\end{eqnarray}
and the integrations over $\hat{R}$ and $\vec{X}$ are performed as 
discussed above.

\section{Analytical expressions for the non-interacting 
and weakly-interacting limits}
\label{sec_appendixc}
This appendix summarizes analytical expressions for the non-interacting 
and weakly-interacting limits.
These results are useful for two reasons. First,
they aid---as illustrated in Sec.~\ref{sec_results}---with the interpretation
of the results for the interacting systems. 
Second, we have used these analytical results to check
our numerical implementations.

We start with the $a_s \rightarrow 0^-$ limit and present
explicit analytical expressions for the one-body density matrix
$\rho_1(\vec{r} \, ',\vec{r})$
and the reduced two-body density matrix $\rho_{\mathrm{red}}(\vec{R} \, ',\vec{R})$,
as well as for quantities derived from $\rho_1(\vec{r} \, ',\vec{r})$
and $\rho_{\mathrm{red}}(\vec{R} \, ',\vec{R})$. 
To illustrate the behavior of these quantities,
we consider the ground state 
of the $(2,2)$ system as an example; other states and other systems can
be treated similarly.  
The ground state wave function of the non-interacting $(2,2)$
atomic Fermi gas has $L^{\Pi}=0^+$ symmetry,
\begin{eqnarray}
\psi_{\mathrm{tot}}(\vec{r}_1, \vec{r}_2,\vec{r}_3,\vec{r}_4)=
\frac{1}{3^{1/2}{\pi}^{3}a_{\mathrm{ho}}^8}
\exp \left( - \sum_{j=1}^4 \frac{\vec{r}_j^2}{2a_{\mathrm{ho}}^2} \right) 
\nonumber \\
\times
(\vec{r}_1 - \vec{r}_2)^T (\vec{r}_3 - \vec{r}_4).
\end{eqnarray}
The spin-up and spin-down atoms both experience
(identical)
non-trivial correlations due
to the anti-symmetrization. 
Applying the definitions of Sec.~\ref{sec_densitymatrix},
we find
\begin{eqnarray}
\label{eq_rho_n4}
\rho_1(\vec{r} \, ',\vec{r})=
\frac{3+2 \vec{r} \, ' \, ^T \vec{r}/a_{\mathrm{ho}}^2}{6{\pi}^{3/2}a_{\mathrm{ho}}^{3}}
\exp \left( -\frac{\vec{r} \, ' \, ^2+\vec{r}^2 }{2a_{\mathrm{ho}}^2} 
\right) ,
\end{eqnarray}
\begin{eqnarray}
\rho_{00}(r',r)=
\frac{1}{2{\pi}^{3/2}a_{\mathrm{ho}}^{3}}
\exp \left( -\frac{\vec{r} \, ' \, ^2+\vec{r}^2 }{2a_{\mathrm{ho}}^2} 
\right),
\end{eqnarray}
and
\begin{eqnarray}
\label{eq_rho_n4proj}
\rho_{10}(r',r)=\rho_{1-1}(r',r)=\rho_{11}(r',r)= \nonumber \\
\frac{r' r}{9{\pi}^{3/2}a_{\mathrm{ho}}^{5}}
\exp \left( -\frac{\vec{r} \, ' \, ^2+\vec{r}^2 }{2a_{\mathrm{ho}}^2} 
\right).
\end{eqnarray}
Higher partial wave projections vanish, i.e.,
$\rho_{lm}(r',r)=0$ for $l>1$.
Diagonalizing the projected one-body density matrices $\rho_{lm}(r',r)$
allows for the determination of the natural orbitals and occupation
numbers. Inspection of Eqs.~(\ref{eq_rho_n4})-(\ref{eq_rho_n4proj})
shows that the one-body density matrix can be decomposed 
into four natural orbitals, 
\begin{eqnarray}
\chi_{000}(\vec{r})=
\frac{1}{{\pi}^{3/4}a_{\mathrm{ho}}^{3/2}}
\exp \left( -\frac{\vec{r}^2 }{2a_{\mathrm{ho}}^2} \right) ,
\end{eqnarray}
\begin{eqnarray}
\chi_{010}(\vec{r})=
\frac{\sqrt{2} z }{{\pi}^{3/4}a_{\mathrm{ho}}^{5/2}}
\exp \left( -\frac{\vec{r}^2 }{2a_{\mathrm{ho}}^2} \right) ,
\end{eqnarray}
and similarly for the $(l,m)=(1,-1)$ and $(1,1)$
components.
The corresponding occupation numbers are
$n_{000}=1/2$ and $n_{010}=n_{01-1}=n_{011}=1/6$, i.e., on average
one of the spin-up
atoms occupies a $(l,m)=(0,0)$ orbital while the second 
spin-up atom occupies a combination of three  $l=1$ 
orbitals. 
For completeness, we also report the expression for
the spherical component $n_{1,\mathrm{sp}}(k)$ of the momentum distribution,
\begin{eqnarray}
\label{eq_momentum_n4onebody}
n_{1,\mathrm{sp}}(k)= \frac{3 a_{\mathrm{ho}}^3+2 a_{\mathrm{ho}}^5 k^2}{6 \pi^{3/2}} \exp(-a_{\mathrm{ho}}^2 k^2).
\end{eqnarray}

Similarly, we analyze the reduced 
two-body density matrix $\rho_{\mathrm{red}}(\vec{R} \, ' ,\vec{R})$.
We find
\begin{eqnarray}
\label{eq_rhored_n4twobody}
\rho_{\mathrm{red}}(\vec{R} \, ',\vec{R})= 
\nonumber \\
\frac{39-12(R \, ' \, ^2+R^2)/a_{\mathrm{ho}}^2+16 R \, ' \, ^2 R^2 /a_{\mathrm{ho}}^4+
16 \vec{R} \, ' \, ^T \vec{R}/a_{\mathrm{ho}}^2}{12\sqrt{2}{\pi}^{3/2}a_{\mathrm{ho}}^{3}} 
\nonumber \\
\times
\exp \left( -\frac{\vec{R} \, ' \, ^2+\vec{R}^2 }{a_{\mathrm{ho}}^2} 
\right) ,
\end{eqnarray}
\begin{eqnarray}
\rho_{00}(R',R)= \nonumber \\
\frac{39-12(R \, ' \, ^2+R^2)/a_{\mathrm{ho}}^2+16R \, ' \, ^2R^2/a_{\mathrm{ho}}^4}
{12 \sqrt{2}{\pi}^{3/2}a_{\mathrm{ho}}^{3}}
\nonumber \\
\times
\exp \left( -\frac{\vec{R} \, ' \, ^2+\vec{R}^2 }{a_{\mathrm{ho}}^2} 
\right),
\end{eqnarray}
and
\begin{eqnarray}
\label{eq_rho_n4projtwobody}
\rho_{10}(R',R)=\rho_{1-1}(R',R)=\rho_{11}(R',R)= \nonumber \\
\frac{2^{3/2} R' R}{9{\pi}^{3/2}a_{\mathrm{ho}}^{5}}
\exp \left( -\frac{\vec{R} \, ' \, ^2+\vec{R}^2 }{a_{\mathrm{ho}}^2} 
\right).
\end{eqnarray}
Higher partial wave projections vanish, i.e.,
$\rho_{lm}(R',R)=0$ for $l>1$.
Diagonalizing the projected reduced two-body density matrices $\rho_{lm}(R',R)$
allows for the determination of the natural orbitals and occupation
numbers. 
Inspection of Eqs.~(\ref{eq_rhored_n4twobody})-(\ref{eq_rho_n4projtwobody})
shows that the 
reduced two-body density matrix can be decomposed 
into five natural orbitals, 
\begin{eqnarray}
\chi_{000}(\vec{R})=
\frac{2^{3/4}}{{\pi}^{3/4}a_{\mathrm{ho}}^{3/2}}
\exp \left( -\frac{\vec{R}^2 }{a_{\mathrm{ho}}^2} \right) ,
\end{eqnarray}
\begin{eqnarray}
\chi_{100}(\vec{R})=
\frac{2^{5/4}(3-4R^2/a_{\mathrm{ho}}^2)}{12^{1/2}{\pi}^{3/4}a_{\mathrm{ho}}^{3/2}}
\exp \left( -\frac{\vec{R}^2 }{a_{\mathrm{ho}}^2} \right) ,
\end{eqnarray}
\begin{eqnarray}
\chi_{010}(\vec{R})=
\frac{2^{7/4} Z }{{\pi}^{3/4}a_{\mathrm{ho}}^{5/2}}
\exp \left( -\frac{\vec{R}^2 }{a_{\mathrm{ho}}^2} \right) ,
\end{eqnarray}
and similarly for the $(l,m)=(1,-1)$ and $(1,1)$
components.
The corresponding occupation numbers are
$N_{000}=5/8$, $N_{100}=1/8$ and $N_{010}=N_{01-1}=N_{011}=1/12$.
For completeness, we also report the expression for
the spherical component $n_{\mathrm{red},\mathrm{sp}}(K)$ of the momentum distribution,
\begin{eqnarray}
\label{eq_momentum_n4red}
n_{\mathrm{red},\mathrm{sp}}(K)= 
\nonumber \\
\frac{39 a_{\mathrm{ho}}^3-2 a_{\mathrm{ho}}^5 K^2+ a_{\mathrm{ho}}^7 K^4}{96 \sqrt{2}\pi^{3/2}} 
\exp\left(- \frac{a_{\mathrm{ho}}^2 K^2}{2} \right).
\end{eqnarray}
Figures~\ref{fig_onebody_occupation} and \ref{fig_twobody_occupation}
in Sec.~\ref{sec_structure2} show the occupation numbers derived 
from the one-body and reduced two-body density matrices
for the $(2,2)$ system as a function of $a_s^{-1}$.
In the $a_s \rightarrow 0^-$ limit,
the results for the
interacting system approach 
the analytical expressions presented here.

Next, we consider the $a_s \rightarrow 0^+$ limit.
Assuming that the spin-balanced 
Fermi system can be described as 
consisting of $N/2$ point bosons of mass $M$, where $M=2m_a$,
the wave function $\psi_{\mathrm{tot}}$ becomes
\begin{eqnarray}
\label{eq_totalboson}
\psi_{\mathrm{tot}}(\vec{R}_1,\cdots,\vec{R}_{N/2})=
\prod_{j=1}^{N/2} \Phi_{\mathrm{boson}}(\vec{R}_j),
\end{eqnarray}
where $\vec{R}_j$ denotes the position vector
of the point boson and $\Phi_{\mathrm{boson}}(\vec{R}_j)$
is the ground state harmonic oscillator orbital,
\begin{eqnarray}
\label{eq_bosonorbital}
\Phi_{\mathrm{boson}}(\vec{R})= 
\frac{1}{{\pi}^{3/4} a_{\mathrm{ho},M}^{3/2}} 
\exp \left( - \frac{\vec{R}^2}{2 a_{\mathrm{ho},M}^2} \right),
\end{eqnarray} 
and $a_{\mathrm{ho},M}=\sqrt{\hbar/(M \omega)}$.
For this system, one
readily finds
$\rho_{\mathrm{boson}}(\vec{R} \, ',\vec{R})=\rho_{\mathrm{boson},\mathrm{sp}}(R',R)=
\Phi_{\mathrm{boson}}^*(\vec{R} \, ')\Phi_{\mathrm{boson}}(\vec{R})$,
$N_{000}^{\mathrm{boson}}=1$, $\chi_{000}^{\mathrm{boson}}(\vec{R})= \Phi_{\mathrm{boson}}(\vec{R})$,
and
\begin{eqnarray}
\label{eq_momentumboson}
n_{\mathrm{boson}}(\vec{K})=n_{\mathrm{boson},\mathrm{sp}}(K)= \nonumber \\
\frac{a_{\mathrm{ho},M}^3}{{\pi}^{3/2}}
\exp \left( - (a_{\mathrm{ho},M} {K})^2 \right).
\end{eqnarray} 
In the $a_s \rightarrow 0^+$ limit, the reduced 
two-body density matrix $\bar{\rho}_{\mathrm{red}}(\vec{R} \, ',\vec{R})$
is expected to approach $\rho_{\mathrm{boson}}(\vec{R} \, ',\vec{R})/N_1$.
The factor of $1/N_1$ arises as follows:
The fermionic system contains $N_1 \times N_2$ 
spin-up---spin-down distances. For any given configuration,
however, only $N_2$ of these distances correspond to a relative
distance vector
of a tightly bound pair in the $a_s \rightarrow 0^+$ limit.
Thus, $\rho_{\mathrm{red}}(\vec{R} \, ',\vec{R})$ can be 
decomposed in the $a_s \rightarrow 0^+$ limit
into two pieces:
The first piece, $\bar{\rho}_{\mathrm{red}}(\vec{R} \, ',\vec{R})$,
accounts for the $N_2$ pairs that are condensed.
The second piece, 
$\rho_{\mathrm{red}}(\vec{R} \, ',\vec{R})-\bar{\rho}_{\mathrm{red}}(\vec{R} \, ',\vec{R})$,
accounts for the $N_2(N_1-1)$ pair distances that belong to
large pairs.
Applying this reasoning, we expect that
the ``second piece'' gives rise to the occupation 
of a large number of natural orbitals,
all with small occupation numbers, while
the ``first piece''
gives rise to the macroscopic occupation of a single
$(l,m)=(0,0)$ natural orbital
[i.e., for the lowest natural orbital
of $\bar{\rho}_{\mathrm{red}}(\vec{R},\vec{R} \, ')$, 
we expect $N_{000}=1/2, 1/2, 1/3$ and $1/3$
for the $(2,1)$, $(2,2)$, $(3,2)$ and $(3,3)$ systems,
respectively].
In summary, we expect 
$\bar{\rho}_{\mathrm{red}}(\vec{R} \, ',\vec{R})=\bar{\rho}_{00}(R',R)=\rho_{00}(R',R)=\rho_{\mathrm{boson}}(R',R)/N_1$
in the $a_s \rightarrow 0^+$ limit.
This is confirmed by our numerical calculations. 

As discussed in Sec.~\ref{sec_densitymatrix}, 
we determine the $\bar{\rho}_{lm}({R}',{R})$
through Metropolis sampling. While this approach 
works in principle, observables determined through this Monte Carlo  
approach are necessarily accompanied by statistical errors; the
reduction of these statistical errors
for non-local observables 
is possible 
but does, in general, require significant computational resources.
In contrast, the $\rho_{lm}(R',R)$ can, in most cases, be determined
quite efficiently within the stochastic variational framework
(see Appendix~\ref{sec_appendixa}).
As shown in Sec.~\ref{sec_structure2}, the quantity ${\rho}_{00}({R}',{R})$
contains valuable information. 

To interpret 
the characteristics of ${\rho}_{00}({R}',{R})$
for finite but small $a_s$, it is useful to consider the internal
structure of the composite bosons, which 
can be described 
approximately 
by assuming that the spin-up and spin-down fermions interact
through a $\delta$-function potential.
In the limit of small $a_s$, the confining potential 
can be neglegted and the internal wave function $\Phi_{\mathrm{int}}(\vec{r}_j)$
of the $j$th
tightly bound
pair becomes
\begin{eqnarray}
\label{eq_bosonpair}
\Phi_{\mathrm{int}}(\vec{r}_j)=
\frac{1}{\sqrt{2 a_s \pi} |\vec{r}_j|} 
\exp \left(- \frac{|\vec{r}_j|}{a_s} \right),
\end{eqnarray}
where $\vec{r}_j$ denotes the distance vector
between the spin-up atom and the spin-down atom 
that form the $j$th composite boson.
Equation~(\ref{eq_bosonpair}) is used to
interpret the peak of $\rho_{00}(R',R)$ that exists at length scales
of the order of $a_s$ (see Fig.~\ref{fig_twobody_diag}).

Lastly, we determine the large $K$ contribution to the
momentum
distribution $n_{\mathrm{red,sp}}(K)$ that 
depends, as discussed in Sec.~\ref{sec_structure2} 
in the context of 
Figs.~\ref{fig_momentumpair} and \ref{fig_momentumpairreferee},
on the internal structure of the molecules.
If $N$ is even, we multiply 
the wave function given in Eq.~(\ref{eq_totalboson})
by
$N/2$ pair functions, i.e., by 
$\prod_{j=1}^{N/2}\Phi_{\mathrm{int}}(\vec{r}_j)$
[see Eq.~(\ref{eq_bosonpair})].
To calculate the large $K$ contribution to 
$n_{\mathrm{red,sp}}(K)$,
we choose 
the $\vec{R}$ and $\vec{R} \, '$
vectors
that enter into $\rho_{red}(\vec{R},\vec{R} \, ')$
to belong to spin-up---spin-down pairs 
that have relatively
large interparticle distances.
For example, if particles 1 and $N/2+1$ form a pair and particles
2 and $N/2+2$ form a pair, then we choose
$\vec{R}=(\vec{r}_1+\vec{r}_{N/2+2})/2$
and 
$\vec{R} \, '=(\vec{r}_2+\vec{r}_{N/2+1})/2$.
Evaluating $\rho_{\mathrm{red}}(\vec{R},\vec{R} \, ')$, 
and in turn $n_{\mathrm{red,sp}}(K)$,
for this choice of coordinates 
and the approximate analytical wave function,
we find 
\begin{eqnarray}
\label{eq_momentummodel}
n_{\mathrm{model,sp}}(K)=
\frac{a_{\mathrm{ho}}^3}{\pi^{3/2} K a_{\mathrm{ho}}} 
\Re \left[ -i A \exp(A^2) \mathrm{Erfc}(A) \right],
\end{eqnarray}
where $A=2 a_{\mathrm{ho}}/a_s + i K a_{\mathrm{ho}}$,
for the contribution to $n_{\mathrm{red,sp}}(K)$ for ``large pairs''.
Combining Eqs.~(\ref{eq_momentumboson}) and (\ref{eq_momentummodel})
and taking into account that systems with $N_1=N_2$
contain, as $a_s$ approaches the $0^+$ limit, 
$N_2$ small and $N_1N_2-N_2$ large pairs,
we obtain Eq.~(\ref{eq_momentumanalytical}) of Sec.~\ref{sec_structure2}.

For systems with $N_1-N_2=1$, the unpaird impurity atom
has to be taken into account. 
Multiplying the wave function
constructed for the fully paired system,
i.e., for $N_1=N_2$, by a single particle
ground state harmonic oscillator wave function for the spare particle
and defining the $\vec{R}$ and $\vec{R} \, '$ vectors in terms of 
the coordinates of the impurity atom and those of
one of the spin-down atoms,
we obtain a third contribution to the momentum
distribution $n_{\mathrm{red,sp}}(K)$ in the $a_s \rightarrow 0^+$ limit,
\begin{eqnarray}
\label{eq_momentumspare}
n_{\mathrm{red,sp}}(K) \approx
\frac{1}{N_1}  
n_{\mathrm{boson,sp}}(K)+ \nonumber \\
\frac{(N_1N_2-2 N_2)}{N_1N_2} 
n_{\mathrm{model,sp}}(K)+ \nonumber \\
\frac{1}{N_1} 
\frac{2a_{\mathrm{ho}}^3}{5\pi^{3/2} K a_{\mathrm{ho}}} 
\Re \left[ -i B \exp(B^2) \mathrm{Erfc}(B) \right],
\end{eqnarray}
where
$B=\sqrt{2/5}(a_{\mathrm{ho}}/a_s + i K a_{\mathrm{ho}})$.
We have checked that Eq.~(\ref{eq_momentumspare}) reproduces
the numerically determined
momentum distributions $n_{\mathrm{red,sp}}(K)$
for the $(2,1)$
and $(3,2)$ systems with small $a_s$, $a_s>0$, well for
$K \lesssim r_0^{-1}$.

\section*{Acknowledgements}
We thank D. Rakshit for checking the 
equations presented in
Appendix~\ref{sec_appendixa}.
Support by the NSF through
grant PHY-0855332
and the ARO
are gratefully acknowledged.


\begin{thebibliography}{10}
\bibitem{footnotenorm}
Throughout this paper, we employ a convention
in which the occupation numbers add up to 1 and not to the
number of particles [see the discussion around
Eqs.~(\ref{eq_nat}) and (\ref{eq_nat_norm})].
\bibitem{footnoteadded}
Although the natural orbitals $\chi_i(\vec{r})$
and occupation 
numbers $n_i$ defined
through Eq.~(\ref{eq_nat}) are characteristic for the spin-up atoms,
the subscript ``$1$'' has been 
suppressed for notational convenience. 
Similarly, the subscript ``$1$''
is suppressed below on the quantities 
$\rho_{lm}(r \, ',r)$ and
$\tilde{\chi}_i(\vec{k})$.
To define the one-body density matrix
for the spin-down atoms, which differs from that for the spin-up atoms
if $N_1-N_2 \ne 0$, one can reorder the particles such that
the first particle is 
a spin-down atom and apply Eqs.~(\ref{eq_onebody})-(\ref{eq_ft})
with ``$1$'' replaced by ``$2$''.
\bibitem{footnote1}
As a result of a mistake in making the plots,
the densities in Fig. 15 of Ref.~\cite{stec08}
are by a factor 2 too large.
In addition, 
to compare the densities of Ref.~\cite{stec08} with those presented here
the different normalizations need to be taken into
account: The radial densities defined in Ref.~\cite{stec08}
are normalized to the number of spin-up and
spin-down atoms as opposed to 1 as done in the present work.
\end{thebibliography}
\end{document}